\documentclass[12pt,preprint]{aastex}
\usepackage{emulateapj5}
\usepackage{apjfonts}
\usepackage{natbib}

\input psfig.tex

\def\aa{{A\&A}}
\def\aas{{ A\&AS}}
\def\aj{{AJ}}
\def\al{$\alpha$}
\def\bet{$\beta$}
\def\amin{$^\prime$}
\def\annrev{{ARA\&A}}
\def\apj{{ApJ}}
\def\apjs{{ApJS}}
\def\asec{$^{\prime\prime}$}
\def\baas{{BAAS}}
\def\cc{cm$^{-3}$}
\def\deg{$^{\circ}$}
\def\ddeg{{\rlap.}$^{\circ}$}
\def\dsec{{\rlap.}$^{\prime\prime}$}
\def\cc{cm$^{-3}$}
\def\e#1{$\times$10$^{#1}$}
\def\etal{{et al. }}
\def\flamb{erg s$^{-1}$ cm$^{-2}$ \AA$^{-1}$}
\def\flux{erg s$^{-1}$ cm$^{-2}$}
\def\fnu{erg s$^{-1}$ cm$^{-2}$ Hz$^{-1}$}
\def\hal{H$\alpha$}
\def\hb{H$\beta$}
\def\hst{{\it HST}}
\def\kms{km s$^{-1}$}
\def\lamb{$\lambda$}
\def\lax{{$\mathrel{\hbox{\rlap{\hbox{\lower4pt\hbox{$\sim$}}}\hbox{$<$}}}$}}
\def\gax{{$\mathrel{\hbox{\rlap{\hbox{\lower4pt\hbox{$\sim$}}}\hbox{$>$}}}$}}
\def\simlt{\lower.5ex\hbox{$\; \buildrel < \over \sim \;$}}
\def\simgt{\lower.5ex\hbox{$\; \buildrel > \over \sim \;$}}
\def\lum{erg s$^{-1}$}
\def\mbh{{$M_{\rm BH}$}}
\def\micron{{$\mu$m}}
\def\mnras{{MNRAS}}
\def\nat{{Nature}}
\def\pasp{{PASP}}
\def\perang{\AA$^{-1}$}
\def\percm2{cm$^{-2}$}
\def\peryr{yr$^{-1}$}
\def\pp{\parshape 2 0truein 6.1truein .3truein 5.5truein}
\def\reference{\noindent\pp}
\def\refindent{\par\noindent\parskip=2pt\hangindent=3pc\hangafter=1 }
\def\solum{$L_\odot$}
\def\solmass{$M_\odot$}
\def\feii{\ion{Fe}{2}}
\def\mgii{\ion{Mg}{2}}
\def\caii{\ion{Ca}{2}}
\def\civ{\ion{C}{4}}
\def\oii{[\ion{O}{2}]}
\def\heii{\ion{He}{2}}
\def\hi{\ion{H}{1}}
\def\hii{\ion{H}{2}}
\def\oiii{[\ion{O}{3}]}
\def\ni{[\ion{N}{1}]}
\def\oi{[\ion{O}{1}]}
\def\nii{[\ion{N}{2}]}
\def\hei{\ion{He}{1}}
\def\sii{[\ion{S}{2}]}
\def\siii{[\ion{S}{3}]}

\def\lhal{$L_{{\rm H}\alpha}$}
\def\lbol{$L_{{\rm bol}}$}
\def\ledd{$L_{{\rm Edd}}$}
\def\sigg{$\sigma_g$}
\def\sigs{$\sigma_*$}
\def\mbh{{$M_{\rm BH}$}}

\slugcomment{To appear in {\it The Astrophysical Journal Supplement Series}.}
\shorttitle{AGN SPECTRAL ATLAS}
\shortauthors{HO \& KIM}

\begin{document}

\title{Magellan Spectroscopy of Low-Redshift Active Galactic Nuclei}

\author{Luis C. Ho\altaffilmark{1} and Minjin Kim\altaffilmark{1,2}}

\altaffiltext{1}{The Observatories of the Carnegie Institution for Science,
813 Santa Barbara Street, Pasadena, CA 91101; lho@obs.carnegiescience.edu}

\altaffiltext{2}{Department of Physics and Astronomy, Frontier Physics
Research Division, Seoul National University, Seoul, Korea; mjkim@astro.snu.ac.kr}

\begin{abstract}
We present an atlas of moderate-resolution ($R \approx 1200-1600$) optical 
spectra of 94 low-redshift ($z$ \lax\ 0.5) active galactic nuclei taken with 
the Magellan 6.5~m Clay Telescope.  The spectra mostly cover the rest-frame 
region $\sim 3600-6000$ \AA.  All the objects have preexisting {\it Hubble 
Space Telescope}\ imaging, and they were chosen as part of an ongoing program 
to investigate the relationship between black hole mass and their host galaxy 
properties.  A significant fraction of the sample has no previous quantitative 
spectroscopic measurements in the literature.  We perform spectral 
decomposition of the spectra and present detailed fits and basic measurements 
of several commonly used broad and narrow emission lines, including 
\oii\ \lamb3727, \heii\ \lamb4686, H\bet, and \oiii\ \lamb\lamb4959, 5007.
Eight of the objects are narrow-line sources that were previously 
misclassified as broad-line (type~1) Seyfert galaxies; of these, five appear 
not to be accretion-powered.
\end{abstract}


\section{Introduction}

The spectral properties of active galactic nuclei (AGNs) are pertinent to many
areas of astrophysics.  With the recent interest in massive black holes and 
their apparently close connection with galaxy evolution (see Ho 2004, and 
references therein), there has been a resurgence of attention on AGN properties
that might lead to expedient methods to estimate black hole masses for large
samples of objects.  A very promising technique exploits the velocity width of 
the broad emission lines in type~1 sources, in combination with the size of 
the line-emitting region estimated from the luminosity of the central source 
(Kaspi et al. 2005; Bentz et al. 2009), to calculate the ``virial mass'' of 
the black hole.  This technique has been calibrated for local AGNs using the 
H\bet\ (Kaspi et al. 2000) and H\al\ (Greene \& Ho 2005b) lines, and for higher 
redshift sources using ultraviolet lines (\civ\ \lamb1549: Vestergaard 2002; 
\mgii\ \lamb2800: McLure \& Jarvis 2002).  At the same time, the 
characteristics of the narrow emission lines themselves provide useful clues 
on the impact of AGN activity on certain aspects of the host galaxy (e.g., 
Netzer et al. 2004; Ho 2005; Kim et al. 2006).  

This paper presents an atlas of moderate-resolution optical spectra of 94
low-redshift ($z$ \lax\ 0.5), mostly broad-line (type~1) AGNs.  The spectra 
have relatively high signal-to-noise ratios (S/N) and moderate resolution ($R = 
\lambda/\Delta\lambda\approx1200-1600$), covering predominantly the rest-frame 
region $\sim 3600-6000$ \AA.  The observations were taken as part of 
an ongoing program to investigate the relationship between active black holes 
and their host galaxies.  The ground-based, optical spectra provide the 
necessary material to estimate black hole masses, and existing images in the 
{\it Hubble Space Telescope (HST)}\ data archives give details on the host 
galaxy morphologies and structural parameters (for initial results, see Kim et 
al. 2007, 2008).  Although many of the AGNs are bright, well-known sources, 
most of them do not have reliable, modern spectra.  Of those that do, the
published spectra often have highly heterogeneous quality or were analyzed in 
a manner inadequate for our purposes.  The majority of the sample, in fact, 
was chosen to overlap with the AGNs selected from the {\it Einstein 
Observatory}\ Extended Medium-Sensitivity Survey (EMSS; Gioia et al.  1990), 
an unbiased subset of which was uniformly surveyed with {\it HST}\ by Schade 
et al. (2000).  Schade et al. objects constitute an important component of
our ongoing host galaxy investigations.  The original optical classifications 
of the EMSS AGNs were based on the spectroscopy of Stocke et al. (1991), but 
these authors did not publish the actual spectra, nor did they present 
quantitative analysis of them.  We have therefore decided to reobserve as many 
of the objects as possible from the list of Schade et al. (2000); of the 76 
objects in their sample, we observed 61 (80\%).  When time permitted, we also 
observed additional targets from the {\it HST}\ studies of low-redshift 
quasars by Hamilton et al. (2002) and Dunlop et al. (2003) because we also 
draw heavily on these samples.  Some of these brighter objects already have 
good-quality spectra in the literature (e.g., Boroson \& Green 1992; Marziani 
et al. 2003), but we reobserved them anyway for the sake of homogeneity.

This paper is organized as follows.  Section 2 describes the observations and
data reductions.  Section 3 presents our method of spectral decomposition 
and the resulting measurements.  The spectral atlas is shown in Section 4.  
Section 5 provides a brief summary.  Distance-dependent quantities are 
calculated assuming the following cosmological parameters: $H_0 = 
71$~\kms~Mpc$^{-1}$, $\Omega_{m} = 0.27$, and $\Omega_{\Lambda} = 0.75$ 
(Spergel \etal\ 2003).

\section{Observations and Data Reductions}

The observations were obtained with the Magellan 6.5~m Clay Telescope on three 
observing runs in 23--28 February 2004, 14--18 September 2004, and 7--10 March 
2005.   Table~1 gives a log of the observations.  The data were acquired as 
part of a backup program for a project that required exceptionally good 
seeing.  We turned to the AGNs whenever the seeing conditions deteriorated to 
\gax\ 1\asec, which is considered relatively poor by the standards of Las 
Campanas Observatory.  

During the two 2004 observing runs, we used the now-retired Boller \& Chivens 
(B\&C) long-slit (length 2\amin) spectrograph equipped with a 2048$\times$515 
Marconi chip.  The 13.5 \micron\ pixels project to a scale of 0\farcs25.  
With a slit width of 0\farcs75, the 

\psfig{file=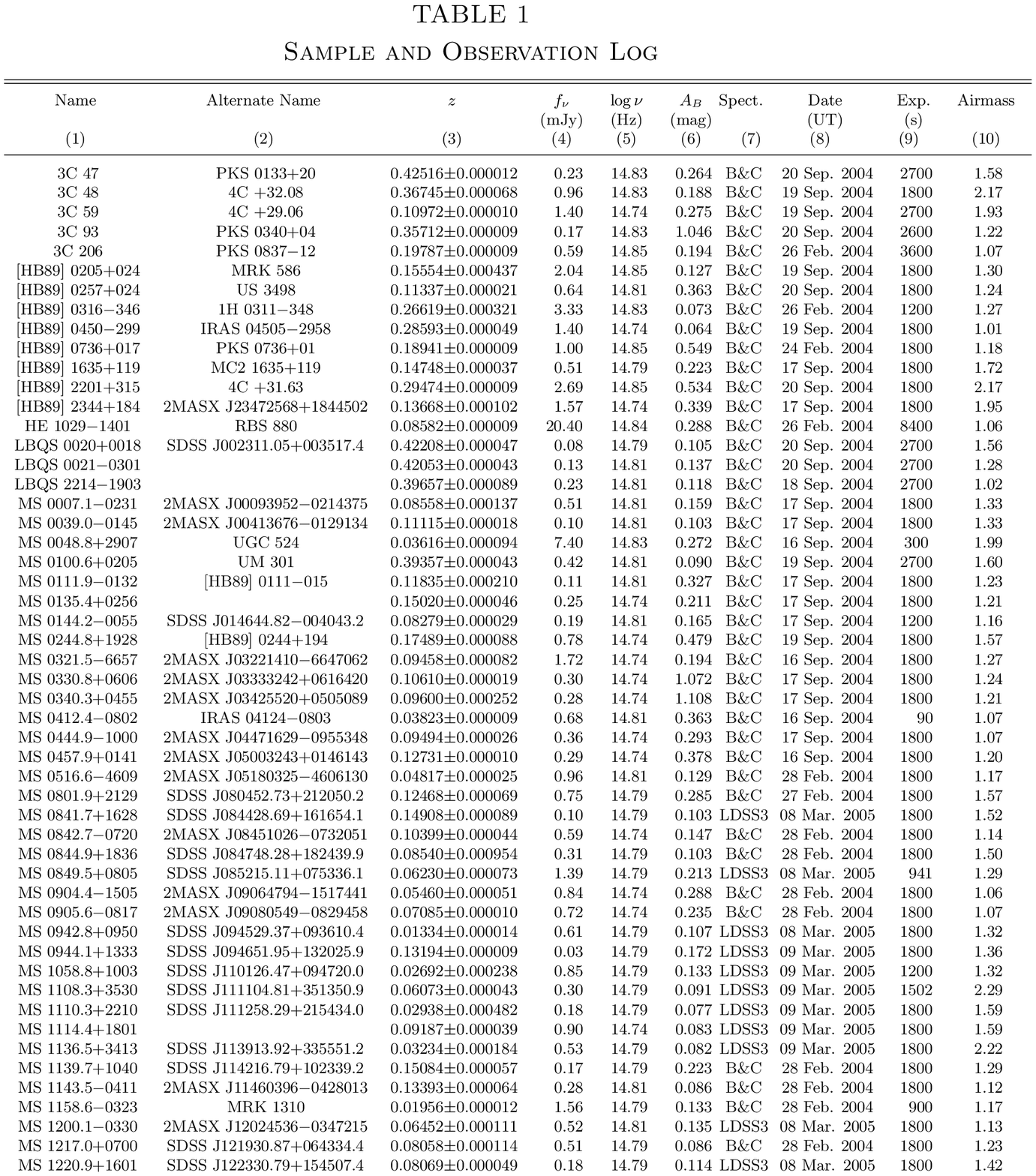,width=18.5cm,angle=0}
\clearpage

\psfig{file=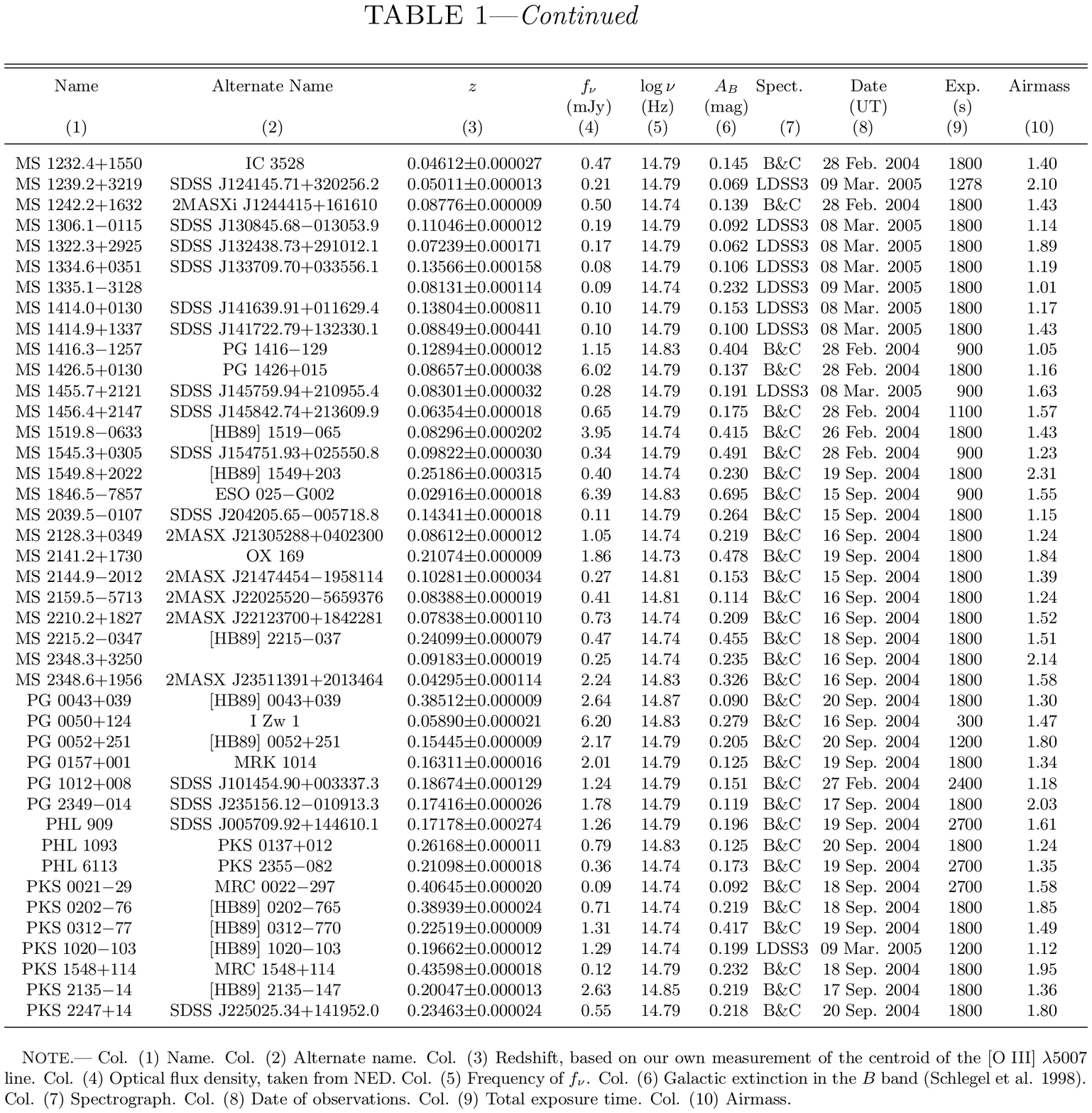,width=18.5cm,angle=0}
\clearpage

\noindent
600 lines mm$^{-1}$ grating gave an average 
full-width at half maximum (FWHM) resolution of 4.2 \AA\ (250 \kms\ at 5000 
\AA).  The spectral resolution, as judged by the widths of the comparison 
arc-lamp spectra and the night sky lines, was relatively uniform across the 
spectrum.  Two grating tilts were used in order to observe the rest-frame 
region of most interest to us, 4200--5750 \AA, which contains the 
diagnostically important lines of \heii\ \lamb4686, H\bet, \oiii\ \lamb\lamb
4959, 5007, and two of the prominent optical \feii\ blends.  For sources with 
$z \le 0.185$, we covered the spectral range $\sim 3640-6820$ \AA; for those 
with $z > 0.185$, the grating tilt was set to cover $\sim 4900-8075$ \AA.  

The 2005 run employed the Low-Dispersion Survey Spectrograph (LDSS3)\footnote{
Information on current Magellan instrumentation can be found at 
{\tt http://www.lco.cl/lco/telescopes-information/magellan/instruments}} in 
long-slit mode.  The 4096$\times$4096 detector has 15 \micron\ pixels and a 
scale of 0\farcs189 pixel$^{-1}$.  We cut three slit masks, each with a 
long slit 0\farcs8 wide, which, in combination with a blue and a red 
volume-phase holographic grating, gave us a total of four spectral 
settings, each covering $\sim 2500$ \AA.  The 1090 lines mm$^{-1}$ blue 
grating has a spectral resolution of FWHM = 3.2 \AA\ (190 \kms\ at 5000 \AA), 
and the 660 lines mm$^{-1}$ red grating has a spectral resolution of FWHM = 
6.5 \AA\ (245 \kms\ at 8000 \AA).

Pixel-to-pixel variations in the response of the CCD were corrected using 
domeflats illuminated by external quartz lamps.  The B\&C data show low-level 
fringing at wavelengths longer than $\sim 6000$ \AA.  To remove this effect, 
for each science image we generated a hybrid flat by combining a 
contemporaneous domeflat for $\lambda > 6000$ \AA\ with a high-S/N flat for 
shorter wavelengths derived from median-combining a large number (40) of 
afternoon domeflats.  Bias correction was achieved by subtracting a constant 
count level determined from the overscan region of the chip.  We took dark 
frames to verify that the CCDs indeed have sufficiently low dark current that 
it can be ignored.  The spectra were wavelength-calibrated using comparison 
arc-lamp spectra, taken at the position of each target, of He+Ar for the B\&C 
runs and He+Ar+Ne for the LDSS3 run.  The wavelength solution is typically 
accurate to $\sim 0.03$ \AA\ rms.

We observed a number of bright G and K giant and subgiant stars, as well as a 
few A-type dwarfs, to model the host galaxy starlight (Section 3).  To perform 
relative flux calibration, we observed spectrophotometric standard stars with 
nearly featureless continua---usually white dwarfs (Stone \& Baldwin 1983; 
Baldwin \& Stone 1984)---at two widely separated airmasses at the beginning 
and end of each night.  Because of the narrowness of the slit and the 
(deliberately) non-optimal conditions of the observations, the absolute flux 
scale is not accurate.  To obtain an approximate absolute flux calibration, we 
empirically bootstrap the observed flux scale to flux densities estimated from 
optical magnitudes collected from the literature (Table~1).  Whenever possible,
we chose literature fluxes that were taken with the smallest possible aperture
in order to mimic our narrow slit width.  As the literature values are quite 
heterogeneous, our final fluxes are only approximate, accurate perhaps to 
no better than a factor of $\sim 2$.

Most of the observations were taken with the slit oriented at the parallactic 
angle to minimize slit losses from atmospheric differential refraction 
(Filippenko 1982).  In a few cases where this was not true, and the airmass 
was substantial, slit losses introduced significant distortion of the continuum 
shape in the blue.  The total exposure time of each target varied from 
90 to 3600~s, depending on the apparent brightness of the source and the 
prevailing weather conditions (some objects were observed under heavy cloud
cover).  To the extent possible, we attempted to reach a uniform minimum S/N 
threshold ($\sim 50-60$ per pixel in the continuum), as judged from real-time, 
quick-look reduction of the data.  This was not always realized because of the 
challenging sky conditions.  Multiple (long $+$ short) exposures were taken of 
some objects to prevent saturation of the brightest emission lines.

We reduced the spectra following standard procedures within the IRAF\footnote{
IRAF is distributed by the National Optical Astronomy Observatory, which is 
operated by the Association of Universities for Research in Astronomy (AURA) 
under cooperative agreement with the National Science Foundation.} package 
{\tt longslit}.  Prior to generating one-dimensional spectra, which uses 
optimal extraction (Horne 1986), we removed cosmic rays using the algorithm of 
van~Dokkum (2001).  Sky subtraction for the B\&C data was performed in a 
straightforward manner by averaging background regions on either side of the 
object.  However, the LDSS3 data required more extensive treatment, 
incorporated into the COSMOS data reduction package\footnote{
{\tt http://www.ociw.edu/Code/cosmos}}, in order to rectify the significant 
curvature in the spatial direction of the images.  In the end, the sky 
subtraction for the LDSS3 data was not fully satisfactory, and regions of the 
spectra near strong night sky lines were adversely affected.  In most 
instances, this has no significant scientific impact, but in a few objects 
important spectral features were partly corrupted.  For cosmetic purposes, in 
the final presentation of the spectra, we removed the small affected regions 
by interpolation.  Finally, regions of the spectrum in long exposures that were 
saturated were replaced with suitably scaled portions from the shorter, 
unsaturated exposure.

Telluric oxygen absorption lines near 6280 \AA, 6860 \AA\ (the ``B band''), 
and 7580 \AA\ (the ``A band'') were removed by division of normalized, 
intrinsically featureless spectra of the standard stars.  Large residuals 
caused by mismatches at the strong bandheads were eliminated by interpolation 
in the plotted spectra.  The reduction procedure also corrected for continuum 
atmospheric extinction.

\section{Spectral Atlas}

Figure 1 presents the spectra for the sources, arranged in increasing 
alphanumeric order.  Each panel shows the final spectrum, corrected for 
foreground Galactic extinction using the $B$-band extinctions given by 
Schlegel et al. (1998) and the extinction curve of Cardelli et al. (1989).  We 
shifted the spectra to the source rest-frame using the heliocentric radial 
velocity determined from the centroid of the narrow core of the \oiii\ \lamb 
5007 line (see Section 4), as listed in Table~1.  The \oiii\ centroid is 
typically accurate to $\sim$0.13 \AA, or $\sim$7.8 \kms, consistent with 
independent checks from published redshifts given in the NASA/IPAC 
Extragalactic Database (NED). For the purposes of the presentation, a few 
objects [whose names are followed by ``(s)''] with particularly low S/N have 
been smoothed with a 5-pixel boxcar.  As explained in Section 2, the flux 
density scale is only approximate; the reader should exercise caution in using 
it for quantitative analysis.  These spectra are available upon request from 
the authors.

\clearpage
%%%%%%%%%%%%%%%%%%%%%%%%%%%%%%%%%%%%%%%%%%%%%%%%%%%%%%%%%%%%%%%%%%%%%%%%%%
\vskip 0.3cm
\begin{figure*}[t]
\figurenum{1}
\centerline{\psfig{file=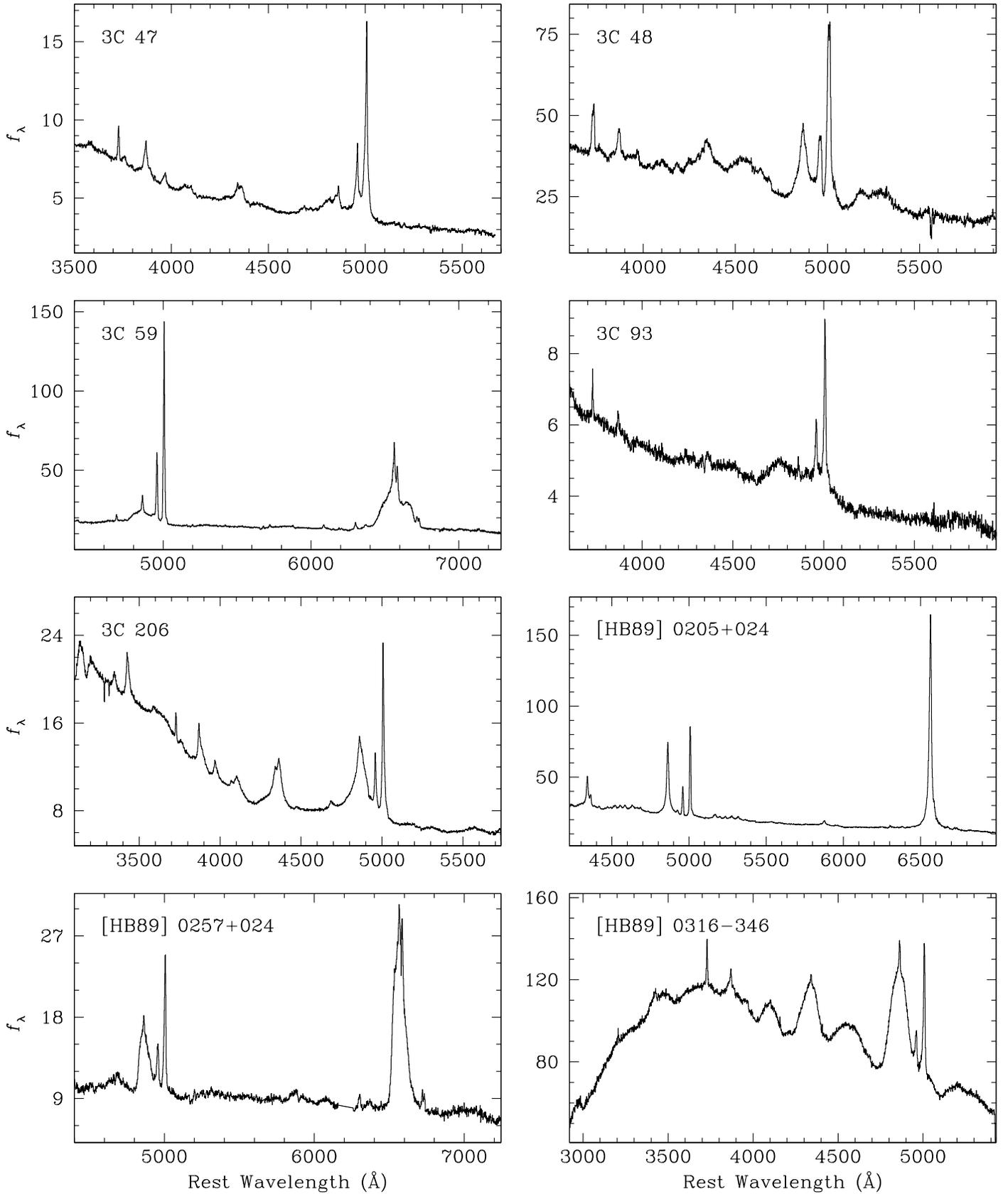,width=19.0cm,angle=0}}
\figcaption[fig1.ps]{Spectral atlas. The ordinate is in units of $10^{-16}$ 
\flamb.  Objects marked with ``(s)'' were smoothed with a 5-pixel boxcar.
\label{fig1}}
\end{figure*}
\vskip 0.3cm
%%%%%%%%%%%%%%%%%%%%%%%%%%%%%%%%%%%%%%%%%%%%%%%%%%%%%%%%%%%%%%%%%%%%%%%%%%%%
\clearpage
%%%%%%%%%%%%%%%%%%%%%%%%%%%%%%%%%%%%%%%%%%%%%%%%%%%%%%%%%%%%%%%%%%%%%%%%%%
\vskip 0.3cm
\begin{figure*}[t]
\figurenum{1}
\centerline{\psfig{file=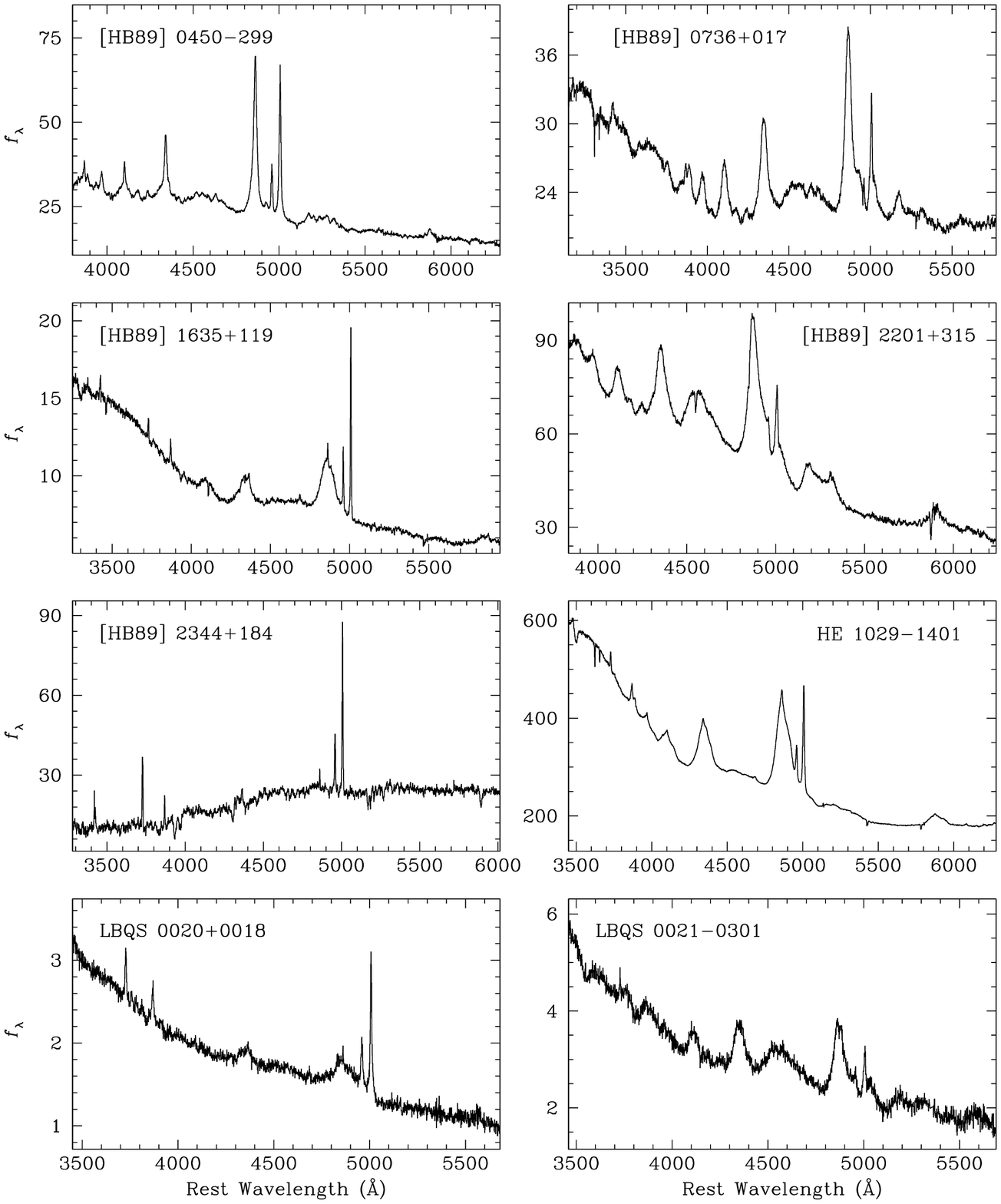,width=19.0cm,angle=0}}
\figcaption[fig1.ps]{Spectral atlas. The ordinate is in units of $10^{-16}$ 
\flamb.  Objects marked with ``(s)'' were smoothed with a 5-pixel boxcar.
\label{fig1}}
\end{figure*}
\vskip 0.3cm
%%%%%%%%%%%%%%%%%%%%%%%%%%%%%%%%%%%%%%%%%%%%%%%%%%%%%%%%%%%%%%%%%%%%%%%%%%%%
\clearpage
%%%%%%%%%%%%%%%%%%%%%%%%%%%%%%%%%%%%%%%%%%%%%%%%%%%%%%%%%%%%%%%%%%%%%%%%%%
\vskip 0.3cm
\begin{figure*}[t]
\figurenum{1}
\centerline{\psfig{file=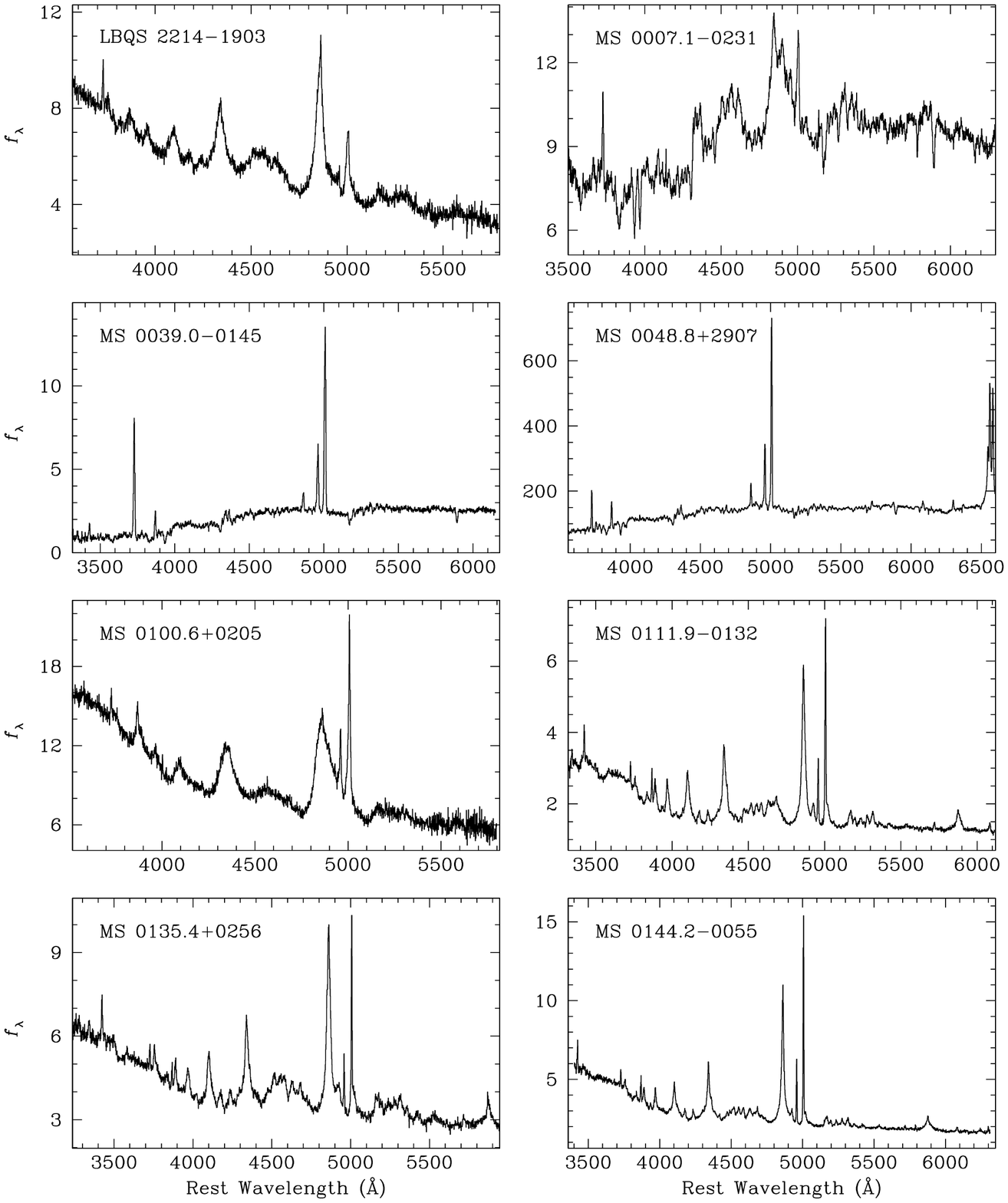,width=19.0cm,angle=0}}
\figcaption[fig1.ps]{Spectral atlas. The ordinate is in units of $10^{-16}$
\flamb.  Objects marked with ``(s)'' were smoothed with a 5-pixel boxcar.
\label{fig1}}
\end{figure*}
\vskip 0.3cm
%%%%%%%%%%%%%%%%%%%%%%%%%%%%%%%%%%%%%%%%%%%%%%%%%%%%%%%%%%%%%%%%%%%%%%%%%%%%
\clearpage
%%%%%%%%%%%%%%%%%%%%%%%%%%%%%%%%%%%%%%%%%%%%%%%%%%%%%%%%%%%%%%%%%%%%%%%%%%
\vskip 0.3cm
\begin{figure*}[t]
\figurenum{1}
\centerline{\psfig{file=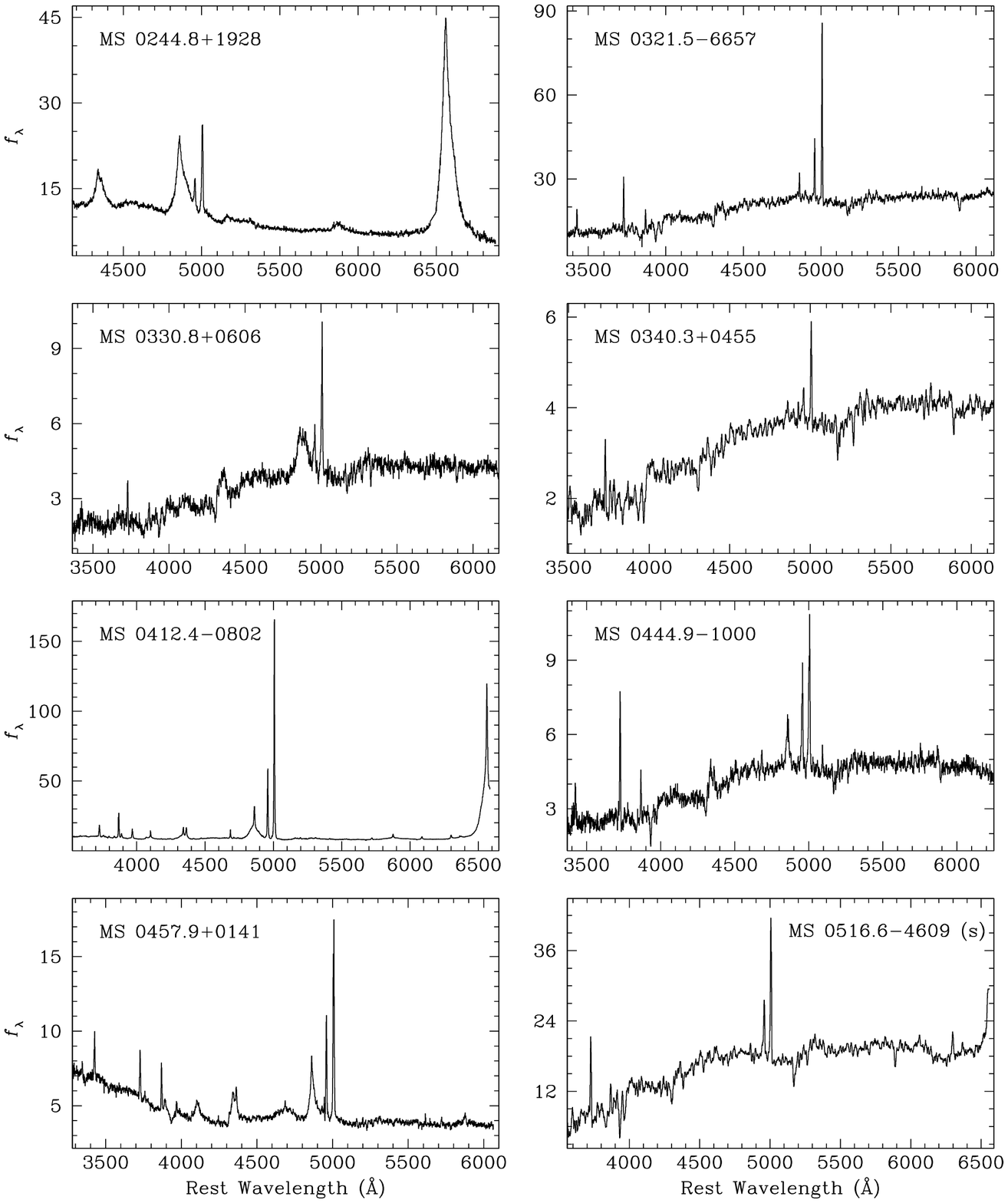,width=19.0cm,angle=0}}
\figcaption[fig1.ps]{Spectral atlas. The ordinate is in units of $10^{-16}$
\flamb.  Objects marked with ``(s)'' were smoothed with a 5-pixel boxcar.
\label{fig1}}
\end{figure*}
\vskip 0.3cm
%%%%%%%%%%%%%%%%%%%%%%%%%%%%%%%%%%%%%%%%%%%%%%%%%%%%%%%%%%%%%%%%%%%%%%%%%%%%
\clearpage
%%%%%%%%%%%%%%%%%%%%%%%%%%%%%%%%%%%%%%%%%%%%%%%%%%%%%%%%%%%%%%%%%%%%%%%%%%
\vskip 0.3cm
\begin{figure*}[t]
\figurenum{1}
\centerline{\psfig{file=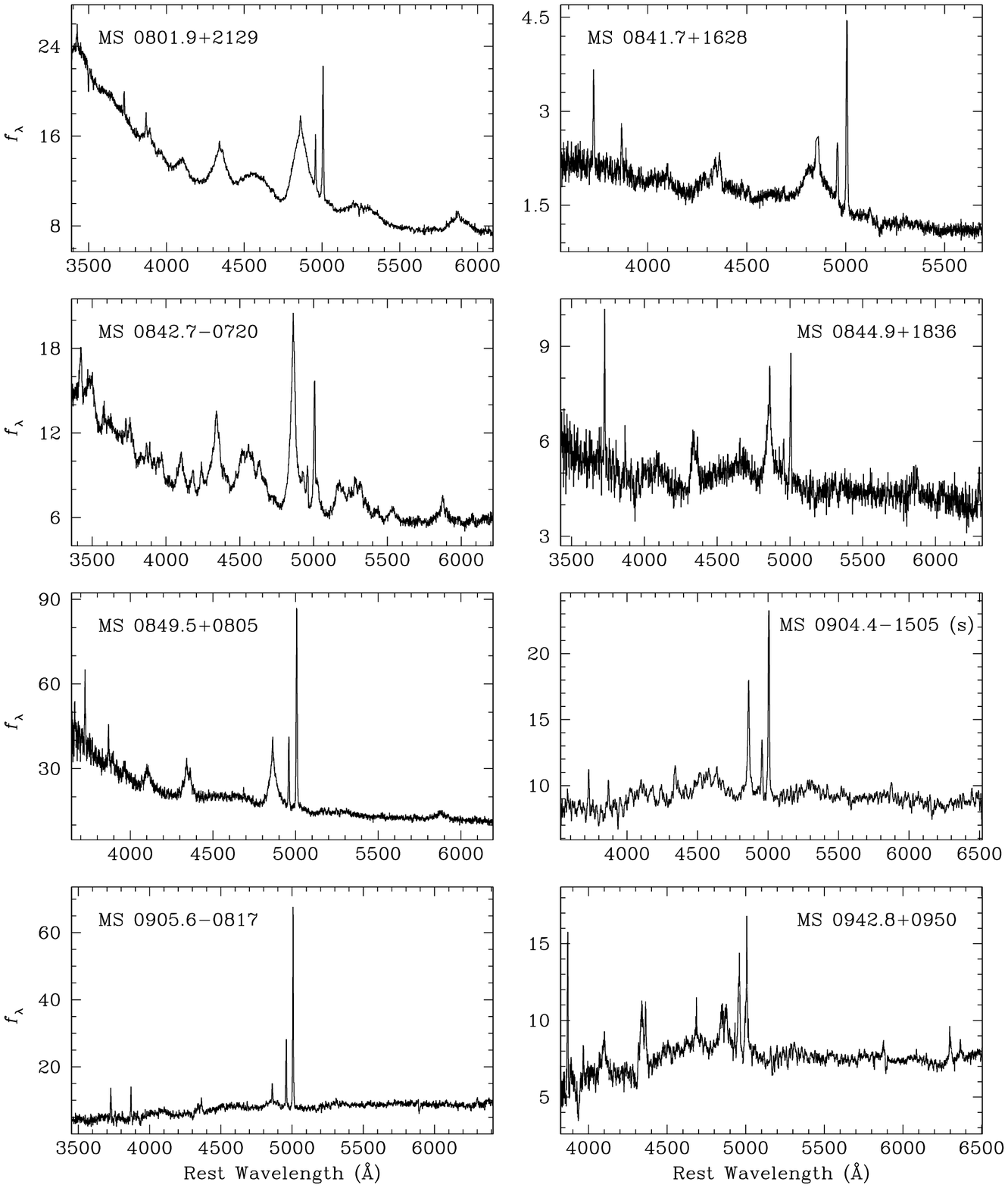,width=19.0cm,angle=0}}
\figcaption[fig1.ps]{Spectral atlas. The ordinate is in units of $10^{-16}$
\flamb.  Objects marked with ``(s)'' were smoothed with a 5-pixel boxcar.
\label{fig1}}
\end{figure*}
\vskip 0.3cm
%%%%%%%%%%%%%%%%%%%%%%%%%%%%%%%%%%%%%%%%%%%%%%%%%%%%%%%%%%%%%%%%%%%%%%%%%%%%
\clearpage
%%%%%%%%%%%%%%%%%%%%%%%%%%%%%%%%%%%%%%%%%%%%%%%%%%%%%%%%%%%%%%%%%%%%%%%%%%
\vskip 0.3cm
\begin{figure*}[t]
\figurenum{1}
\centerline{\psfig{file=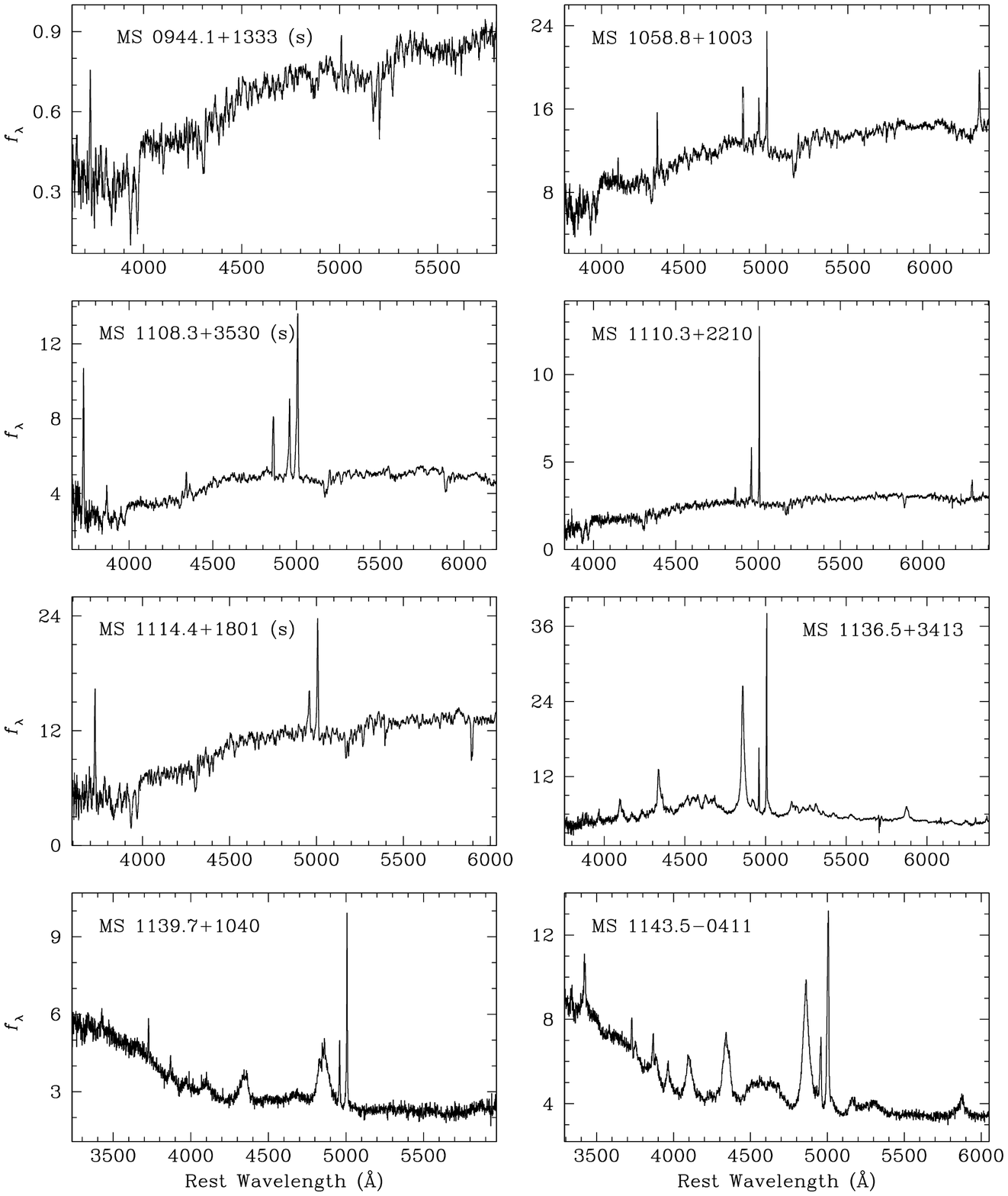,width=19.0cm,angle=0}}
\figcaption[fig1.ps]{Spectral atlas. The ordinate is in units of $10^{-16}$
\flamb.  Objects marked with ``(s)'' were smoothed with a 5-pixel boxcar.
\label{fig1}}
\end{figure*}
\vskip 0.3cm
%%%%%%%%%%%%%%%%%%%%%%%%%%%%%%%%%%%%%%%%%%%%%%%%%%%%%%%%%%%%%%%%%%%%%%%%%%%%
\clearpage
%%%%%%%%%%%%%%%%%%%%%%%%%%%%%%%%%%%%%%%%%%%%%%%%%%%%%%%%%%%%%%%%%%%%%%%%%%%
\vskip 0.3cm
\begin{figure*}[t]
\figurenum{1}
\centerline{\psfig{file=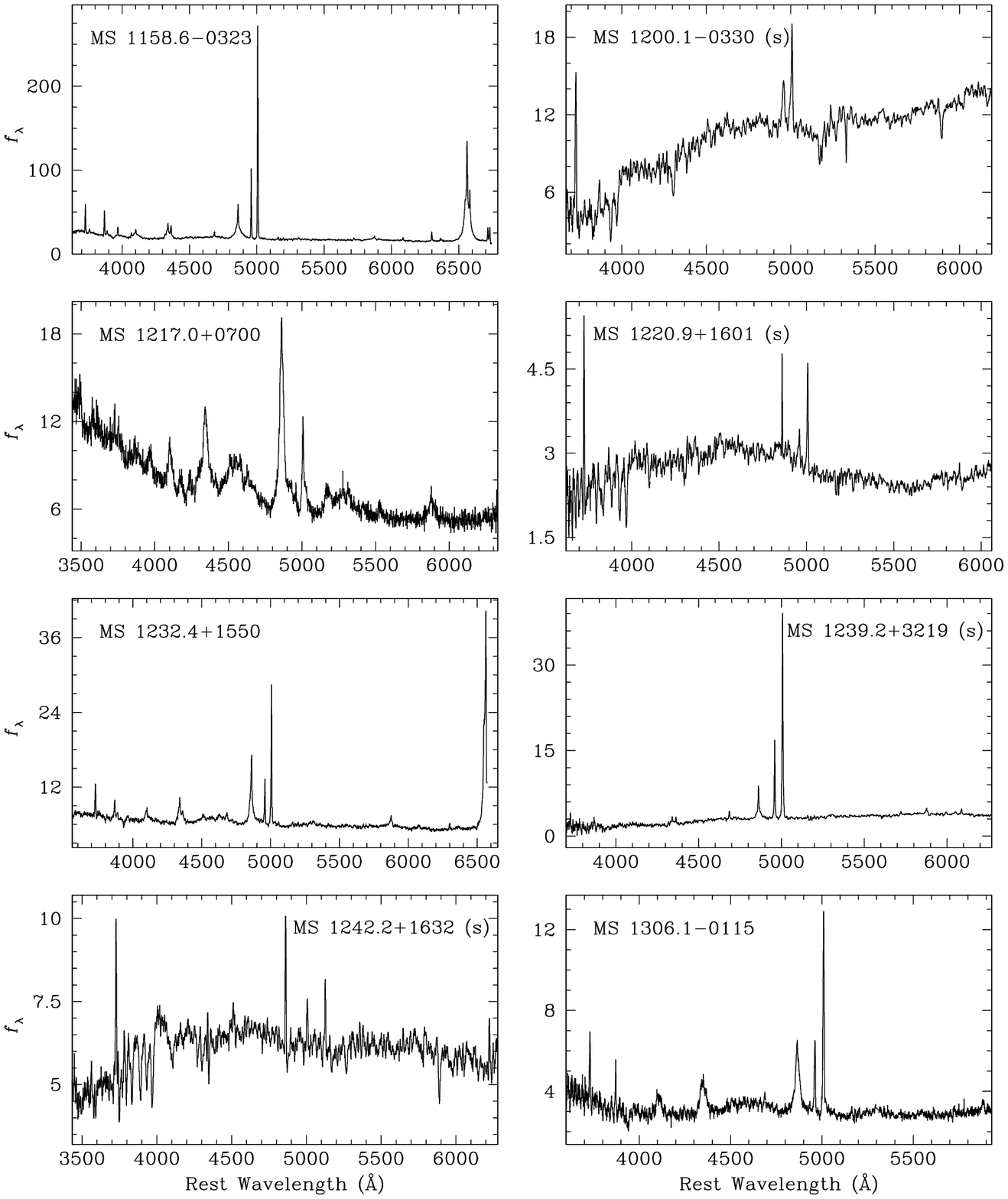,width=19.0cm,angle=0}}
\figcaption[fig1.ps]{Spectral atlas. The ordinate is in units of $10^{-16}$
\flamb.  Objects marked with ``(s)'' were smoothed with a 5-pixel boxcar.
\label{fig1}}
\end{figure*}
\vskip 0.3cm
%%%%%%%%%%%%%%%%%%%%%%%%%%%%%%%%%%%%%%%%%%%%%%%%%%%%%%%%%%%%%%%%%%%%%%%%%%%%
\clearpage
%%%%%%%%%%%%%%%%%%%%%%%%%%%%%%%%%%%%%%%%%%%%%%%%%%%%%%%%%%%%%%%%%%%%%%%%%%
\vskip 0.3cm
\begin{figure*}[t]
\figurenum{1}
\centerline{\psfig{file=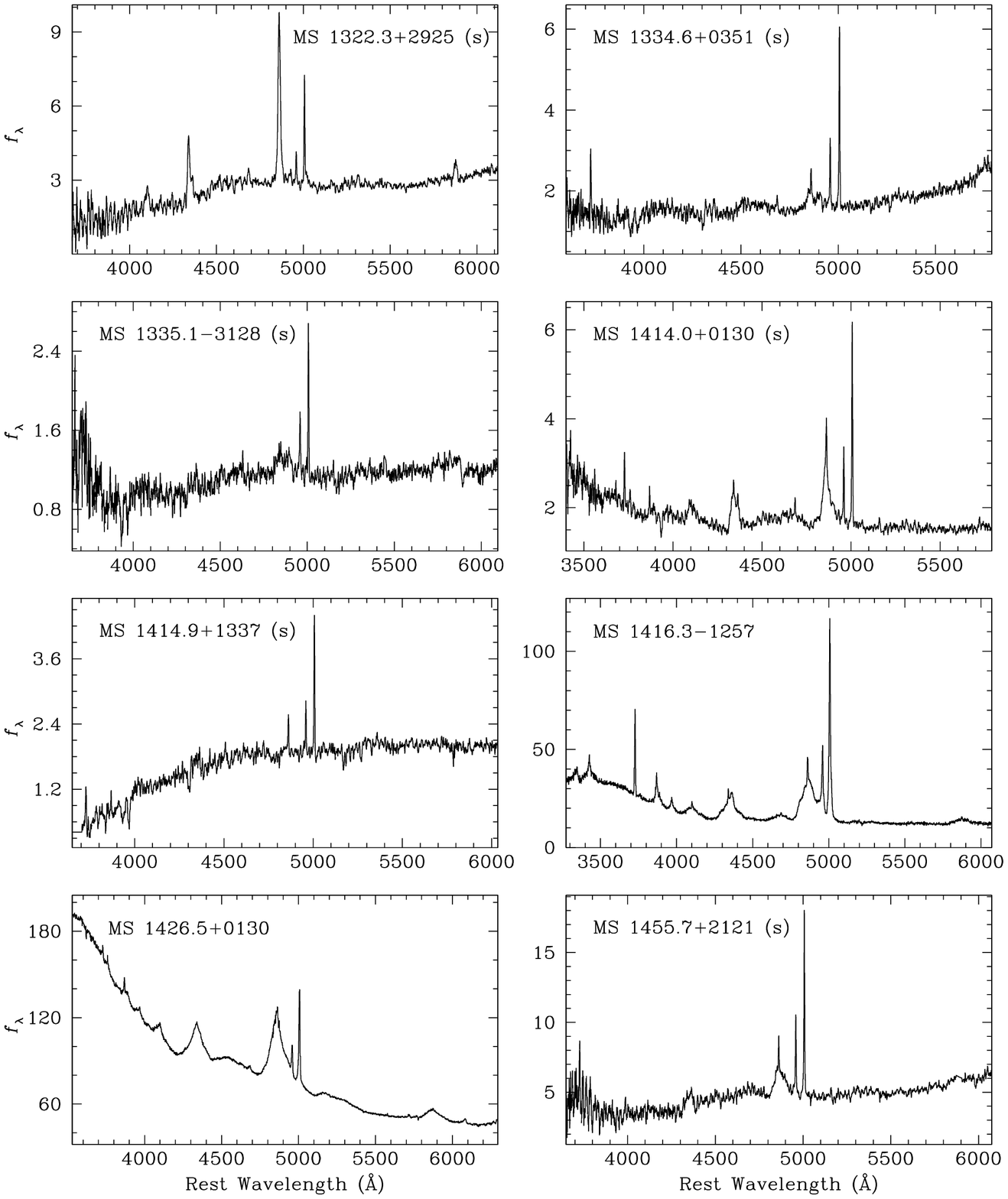,width=19.0cm,angle=0}}
\figcaption[fig1.ps]{Spectral atlas. The ordinate is in units of $10^{-16}$
\flamb.  Objects marked with ``(s)'' were smoothed with a 5-pixel boxcar.
\label{fig1}}
\end{figure*}
\vskip 0.3cm
%%%%%%%%%%%%%%%%%%%%%%%%%%%%%%%%%%%%%%%%%%%%%%%%%%%%%%%%%%%%%%%%%%%%%%%%%%%%
\clearpage
%%%%%%%%%%%%%%%%%%%%%%%%%%%%%%%%%%%%%%%%%%%%%%%%%%%%%%%%%%%%%%%%%%%%%%%%%%
\vskip 0.3cm
\begin{figure*}[t]
\figurenum{1}
\centerline{\psfig{file=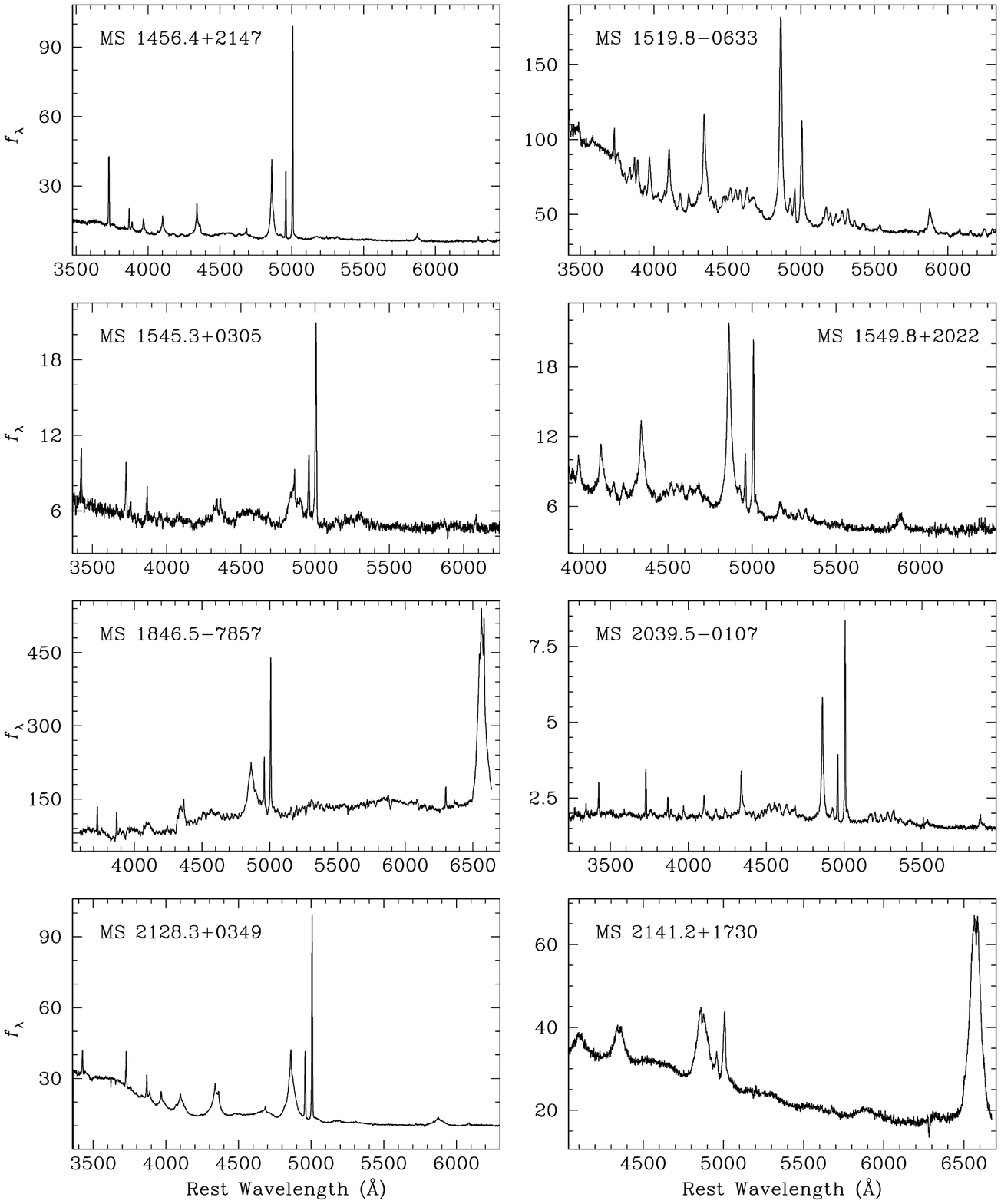,width=19.0cm,angle=0}}
\figcaption[fig1.ps]{Spectral atlas. The ordinate is in units of $10^{-16}$
\flamb.  Objects marked with ``(s)'' were smoothed with a 5-pixel boxcar.
\label{fig1}}
\end{figure*}
\vskip 0.3cm
%%%%%%%%%%%%%%%%%%%%%%%%%%%%%%%%%%%%%%%%%%%%%%%%%%%%%%%%%%%%%%%%%%%%%%%%%%%%
\clearpage
%%%%%%%%%%%%%%%%%%%%%%%%%%%%%%%%%%%%%%%%%%%%%%%%%%%%%%%%%%%%%%%%%%%%%%%%%%
\vskip 0.3cm
\begin{figure*}[t]
\figurenum{1}
\centerline{\psfig{file=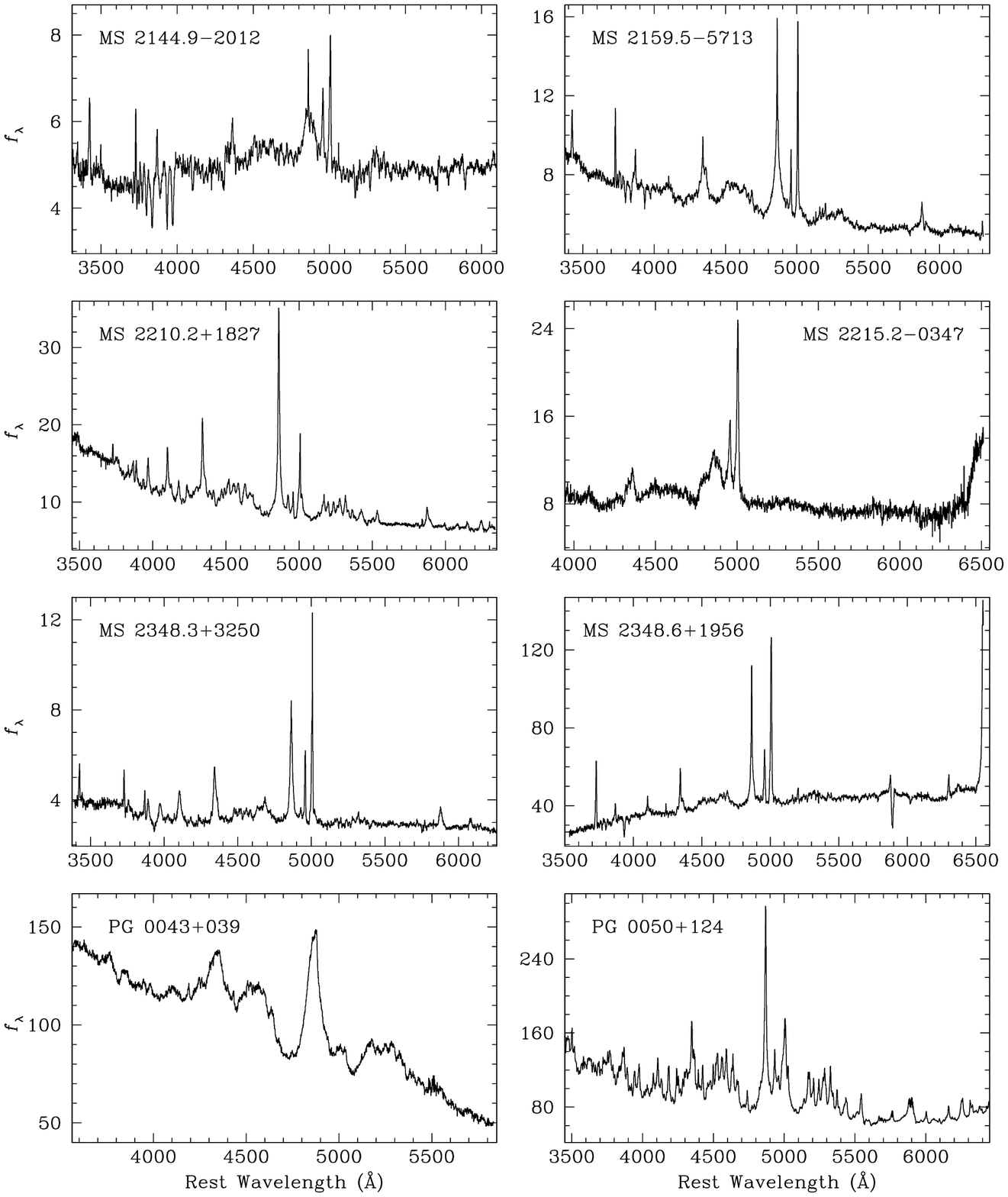,width=19.0cm,angle=0}}
\figcaption[fig1.ps]{Spectral atlas. The ordinate is in units of $10^{-16}$
\flamb.  Objects marked with ``(s)'' were smoothed with a 5-pixel boxcar.
\label{fig1}}
\end{figure*}
\vskip 0.3cm
%%%%%%%%%%%%%%%%%%%%%%%%%%%%%%%%%%%%%%%%%%%%%%%%%%%%%%%%%%%%%%%%%%%%%%%%%%%%
\clearpage
%%%%%%%%%%%%%%%%%%%%%%%%%%%%%%%%%%%%%%%%%%%%%%%%%%%%%%%%%%%%%%%%%%%%%%%%%%
\vskip 0.3cm
\begin{figure*}[t]
\figurenum{1}
\centerline{\psfig{file=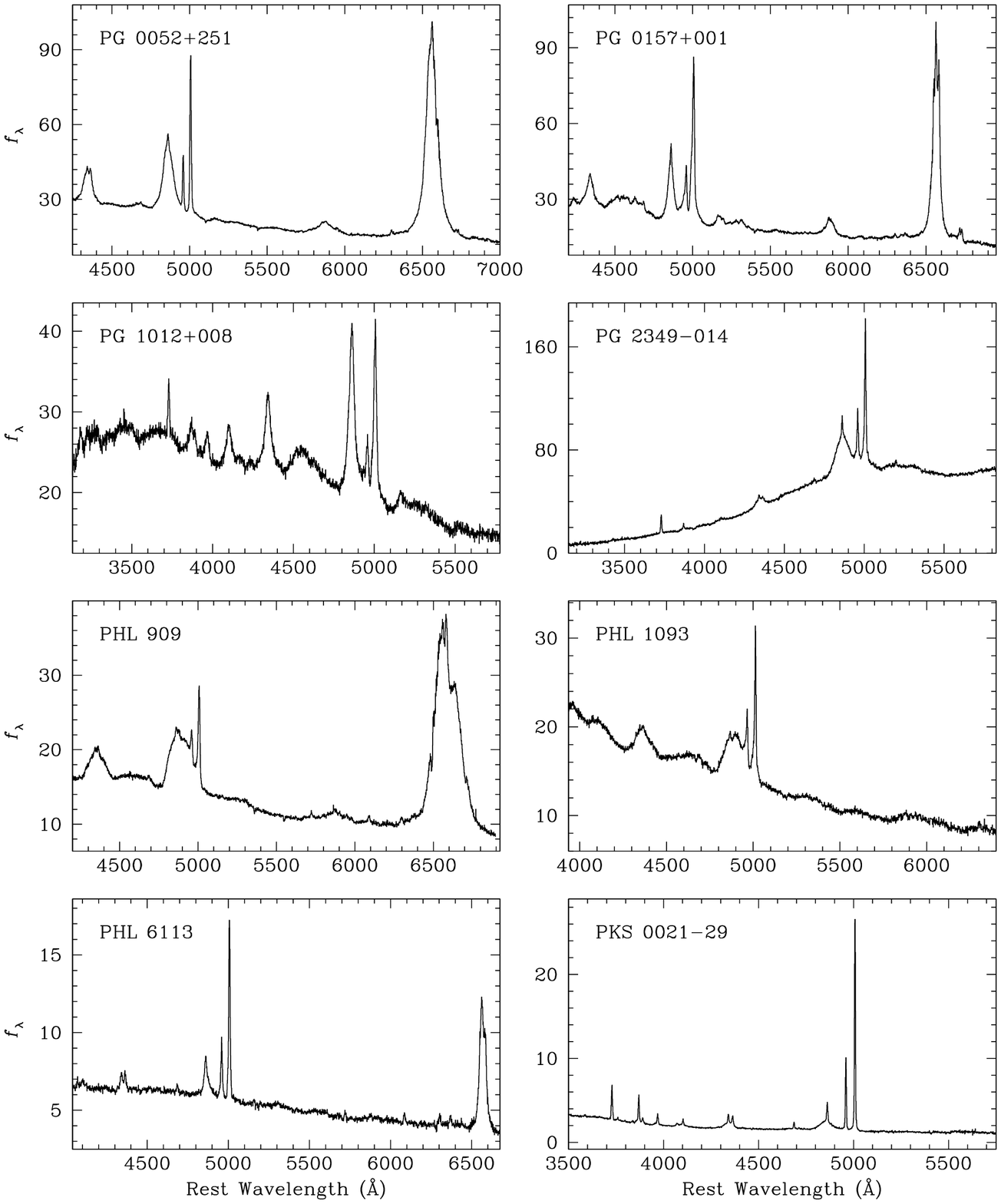,width=19.0cm,angle=0}}
\figcaption[fig1.ps]{Spectral atlas. The ordinate is in units of $10^{-16}$
\flamb.  Objects marked with ``(s)'' were smoothed with a 5-pixel boxcar.
\label{fig1}}
\end{figure*}
\vskip 0.3cm
%%%%%%%%%%%%%%%%%%%%%%%%%%%%%%%%%%%%%%%%%%%%%%%%%%%%%%%%%%%%%%%%%%%%%%%%%%%%
\clearpage
%%%%%%%%%%%%%%%%%%%%%%%%%%%%%%%%%%%%%%%%%%%%%%%%%%%%%%%%%%%%%%%%%%%%%%%%%%
\vskip 0.3cm
\begin{figure*}[t]
\figurenum{1}
\centerline{\psfig{file=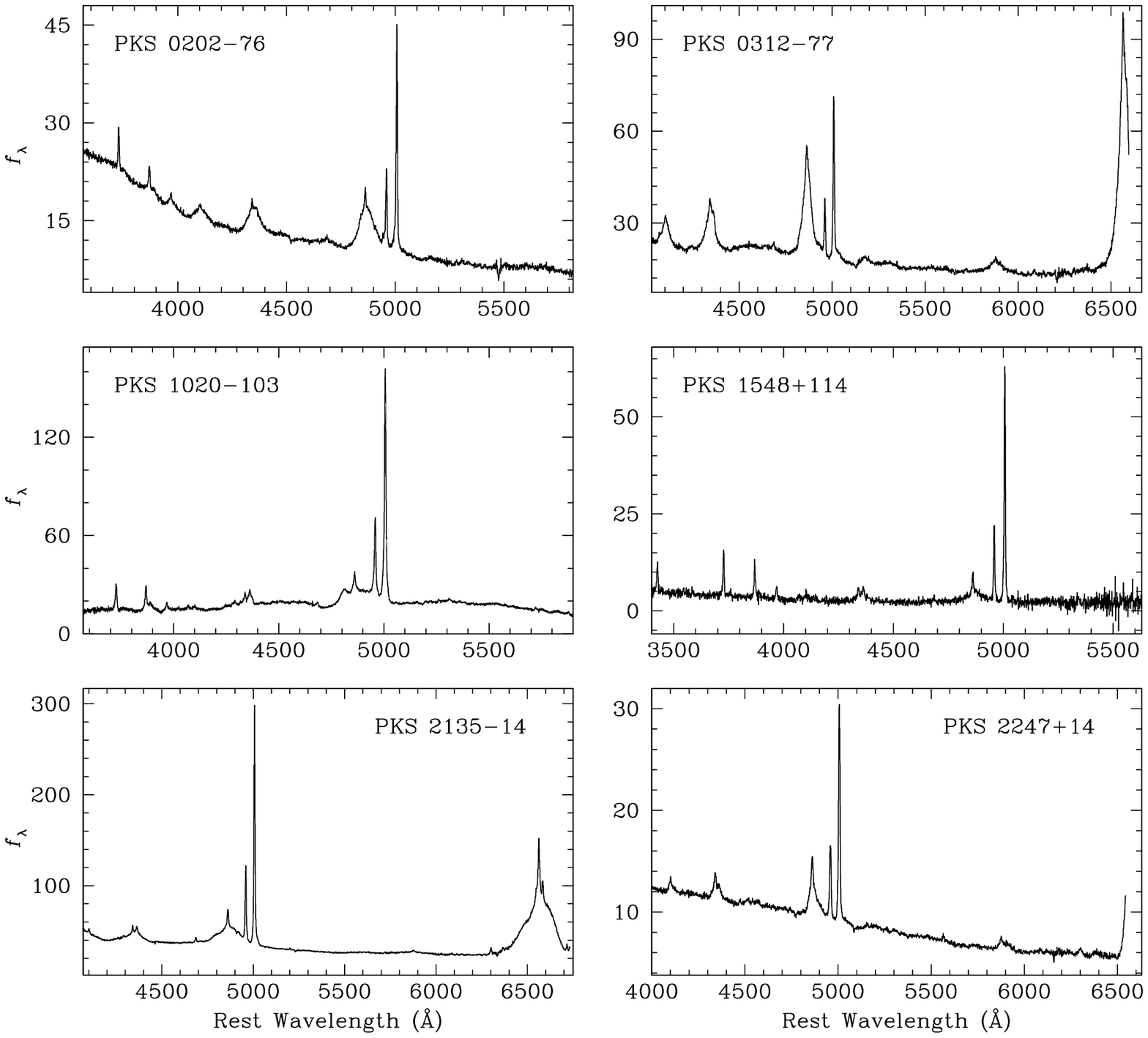,width=19.0cm,angle=0}}
\figcaption[fig1.ps]{Spectral atlas. The ordinate is in units of $10^{-16}$
\flamb.  Objects marked with ``(s)'' were smoothed with a 5-pixel boxcar.
\label{fig1}}
\end{figure*}
\vskip 0.3cm
%%%%%%%%%%%%%%%%%%%%%%%%%%%%%%%%%%%%%%%%%%%%%%%%%%%%%%%%%%%%%%%%%%%%%%%%%%%%
\clearpage

%%%%%%%%%%%%%%%%%%%%%%%%%%%%%%%%%%%%%%%%%%%%%%%%%%%%%%%%%%%%%%%%%%%%%%%%%%
\vskip 0.3cm
\begin{figure*}[t]
\figurenum{2}
\centerline{\psfig{file=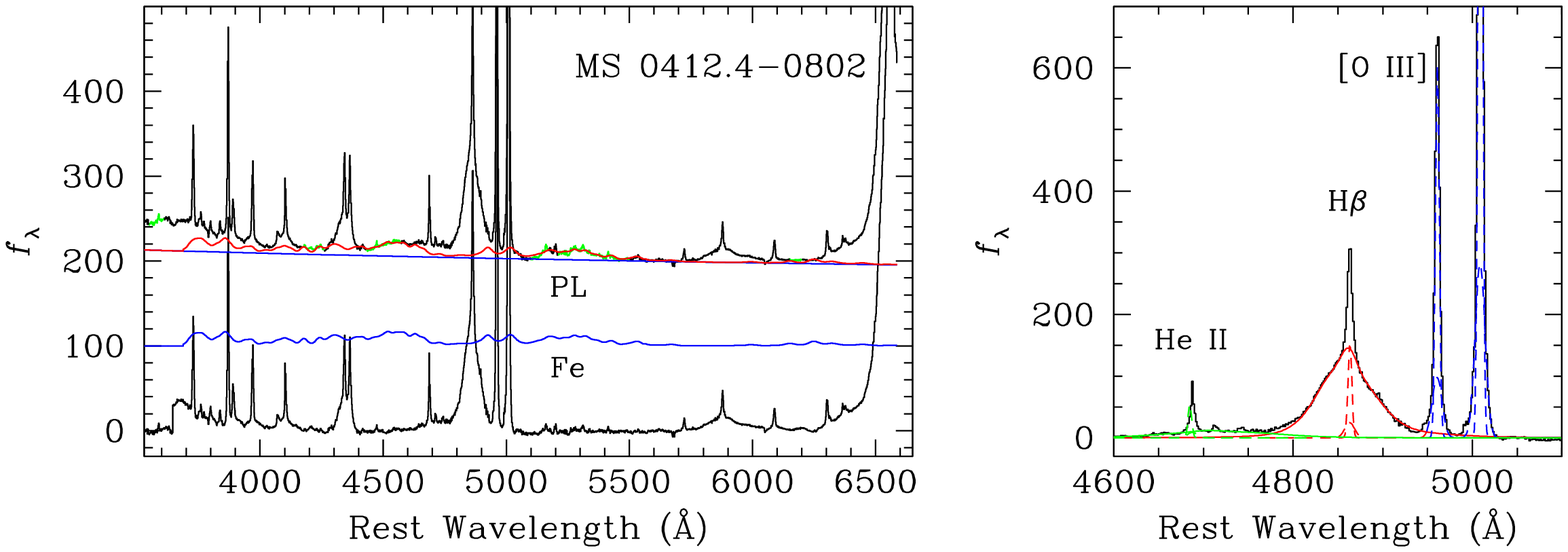,width=19.0cm,angle=0}}
\figcaption[fig2.ps]{Example of spectral decomposition for MS 0412.4$-$0802. 
The ordinate is in units of $10^{-17}$ \flamb. The {\it left}\ panel shows 
the original data ({\it black histograms}), the different components of the 
continuum fit ({\it blue lines}; ``PL'' = power law, ``Fe'' = iron template), 
and the final model ({\it red line}), each offset in the ordinate by a 
constant, arbitrary amount for clarity.  The regions used in the fit are 
highlighted in {\it green}.  The bottom plot shows the residual, pure 
emission-line spectrum.  The {\it right}\ panel illustrates our procedure for 
line fitting for He~{\scriptsize II} ({\it green}), H\bet\ ({\it red}), and 
[O~~{\scriptsize III}] \lamb\lamb4959, 5007 ({\it blue}); solid and dashed
lines denote broad and narrow components, respectively.  
\label{fig2}}
\end{figure*}
\vskip 0.3cm
%%%%%%%%%%%%%%%%%%%%%%%%%%%%%%%%%%%%%%%%%%%%%%%%%%%%%%%%%%%%%%%%%%%%%%%%%%%%

%%%%%%%%%%%%%%%%%%%%%%%%%%%%%%%%%%%%%%%%%%%%%%%%%%%%%%%%%%%%%%%%%%%%%%%%%%
\vskip 0.3cm
\begin{figure*}[t]
\figurenum{3}
\centerline{\psfig{file=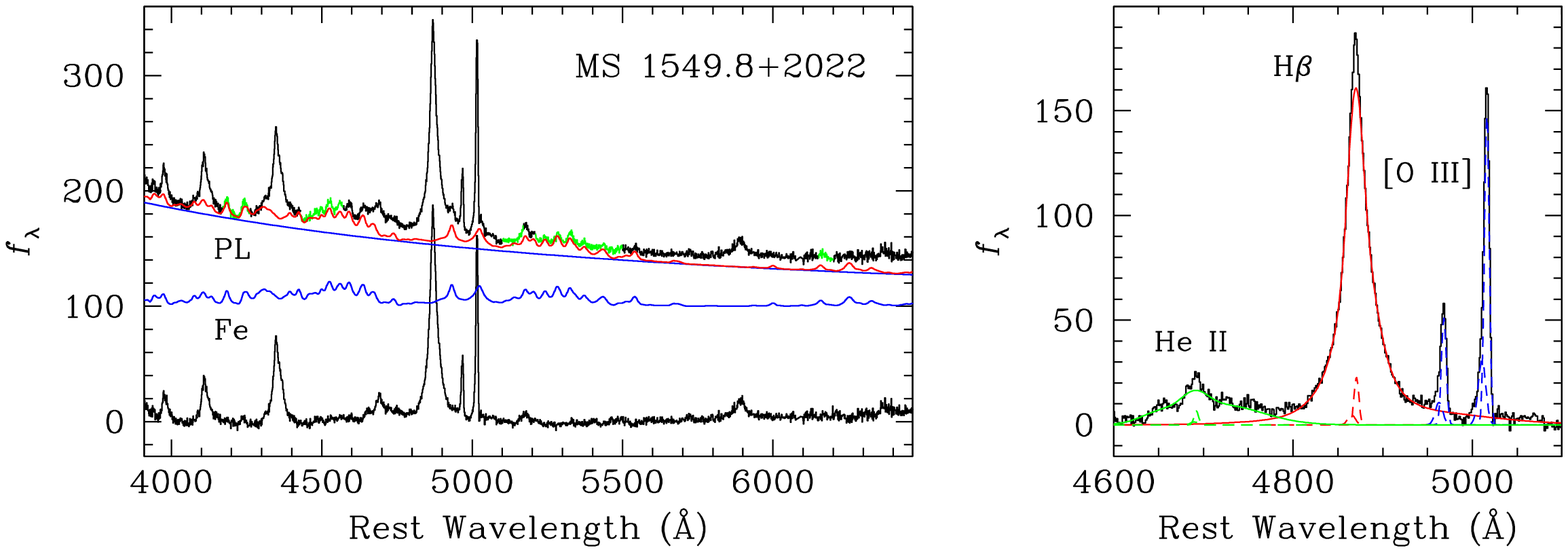,width=19.0cm,angle=0}}
\figcaption[fig3.ps]{Example of spectral decomposition for MS 1549.8$+$2022; 
conventions same as in Figure~2.
\label{fig3}}
\end{figure*}
\vskip 0.3cm
%%%%%%%%%%%%%%%%%%%%%%%%%%%%%%%%%%%%%%%%%%%%%%%%%%%%%%%%%%%%%%%%%%%%%%%%%%%%
\clearpage

%%%%%%%%%%%%%%%%%%%%%%%%%%%%%%%%%%%%%%%%%%%%%%%%%%%%%%%%%%%%%%%%%%%%%%%%%%
\vskip 0.3cm
\begin{figure*}[t]
\figurenum{4}
\centerline{\psfig{file=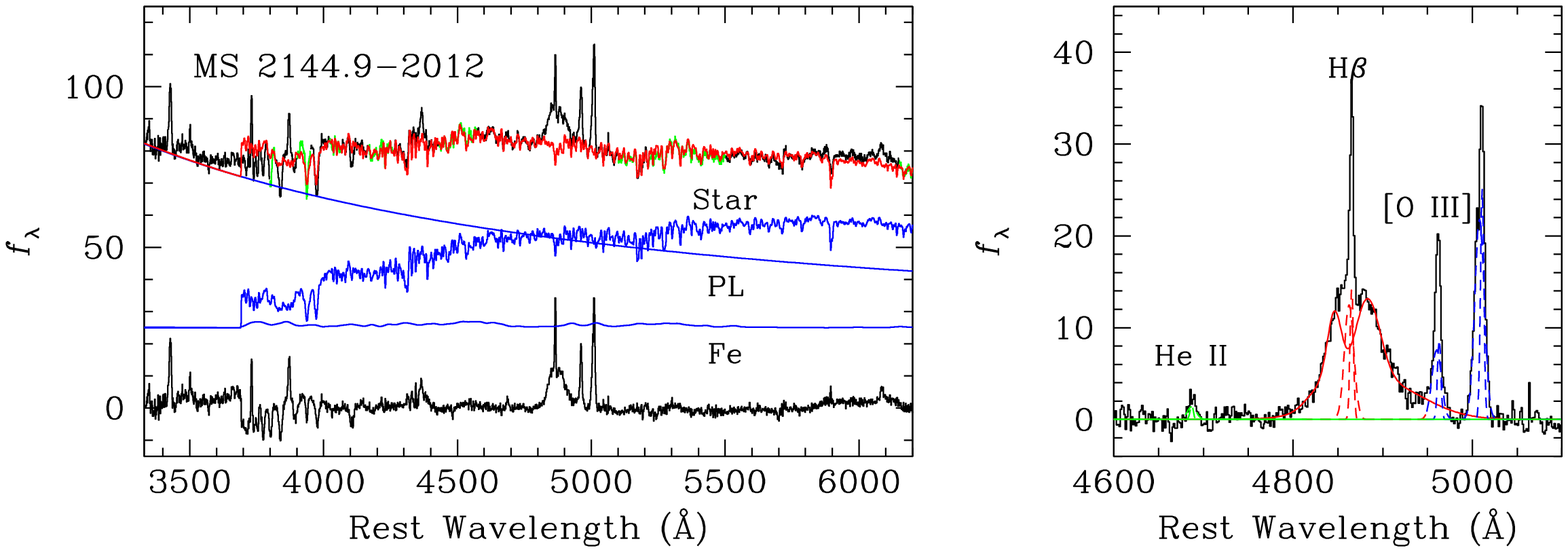,width=19.0cm,angle=0}}
\figcaption[fig4.ps]{Example of spectral decomposition for MS 2144.9$-$2012;
conventions same as in Figure~2, except that the continuum model in this case 
contains a component for starlight (``Star'') from the host galaxy.
\label{fig4}}
\end{figure*}
\vskip 0.3cm
%%%%%%%%%%%%%%%%%%%%%%%%%%%%%%%%%%%%%%%%%%%%%%%%%%%%%%%%%%%%%%%%%%%%%%%%%%%%

\clearpage

\hskip -4in
\psfig{file=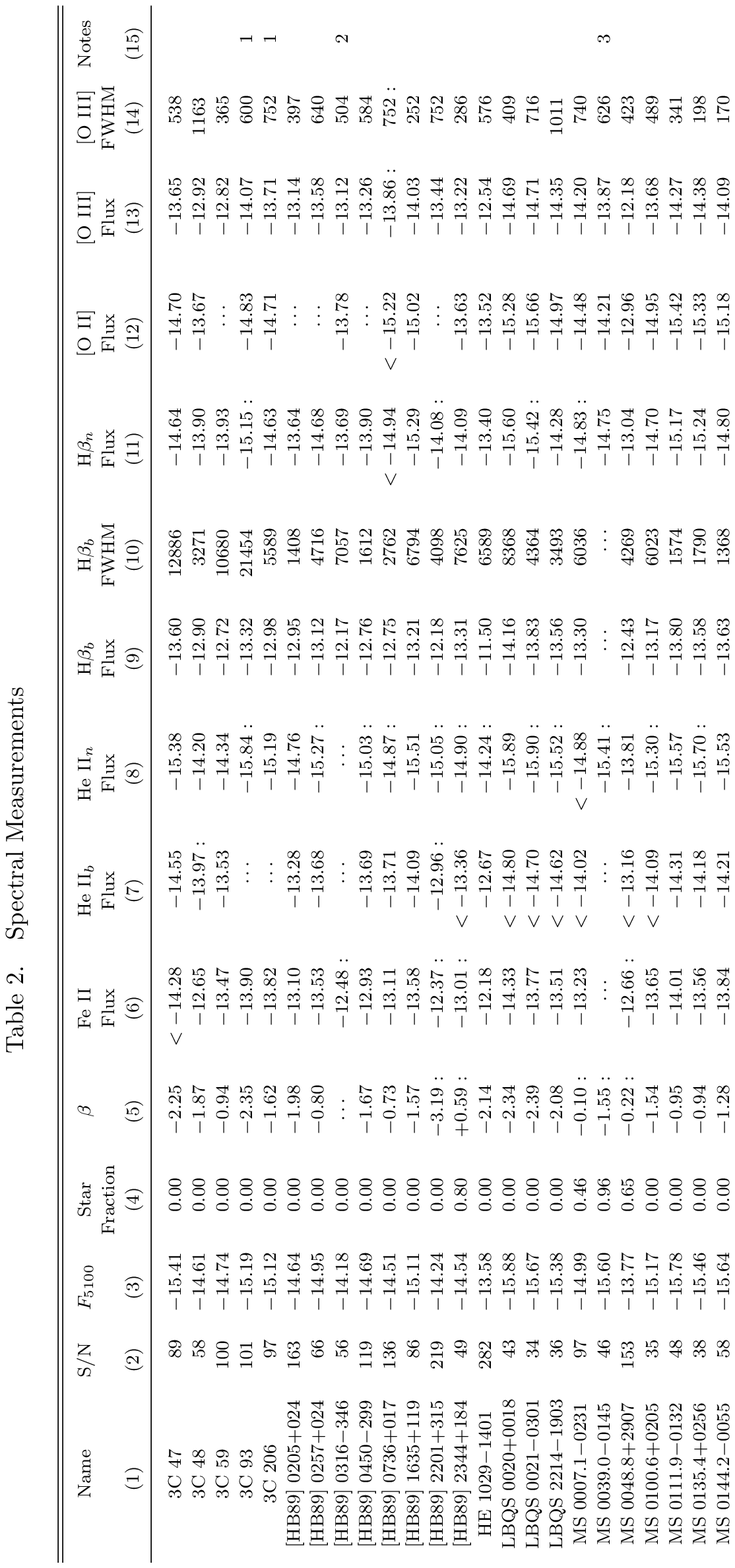,width=7.8in,angle=0}
\clearpage

\hskip -4in
\psfig{file=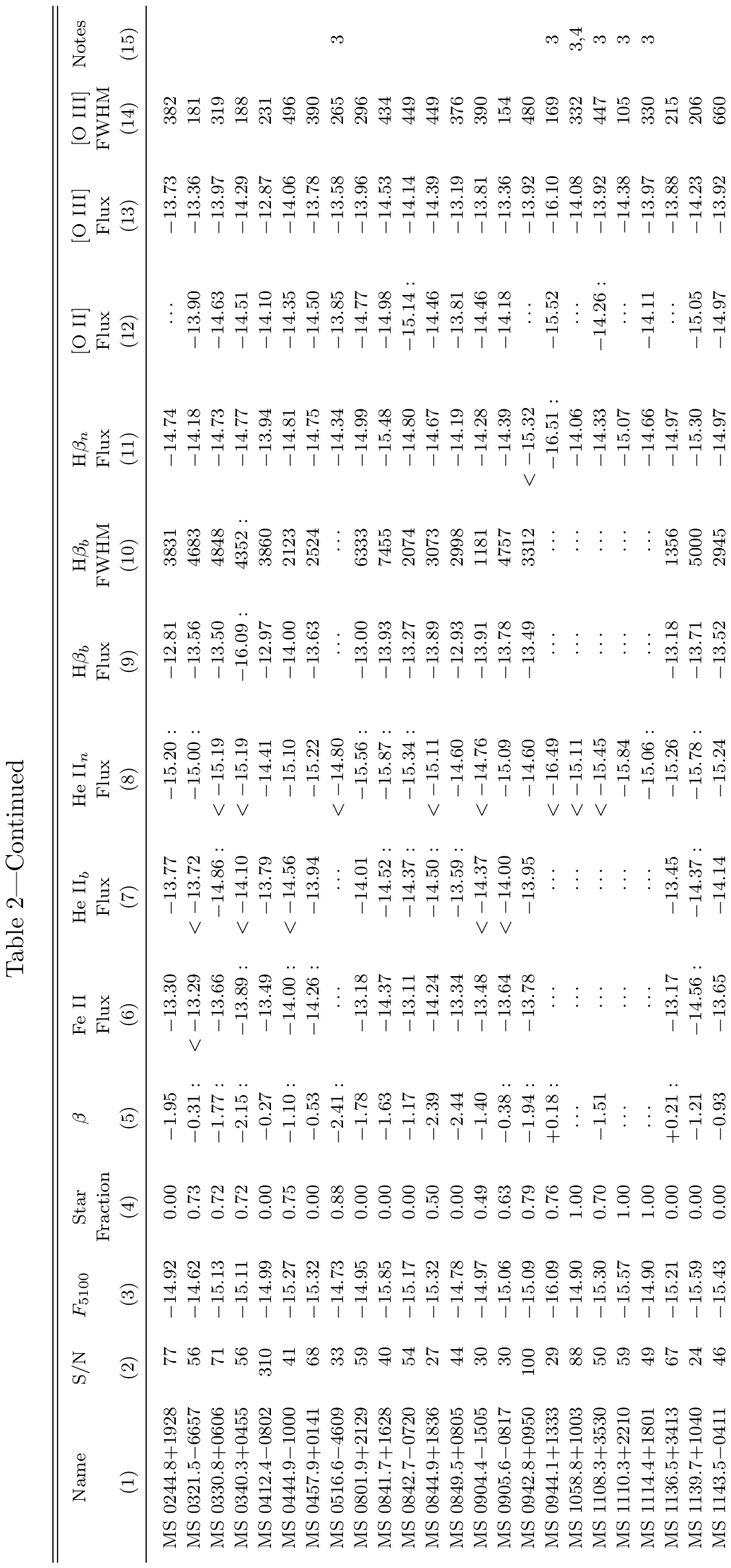,width=7.8in,angle=0}
\clearpage

\hskip -4in
\psfig{file=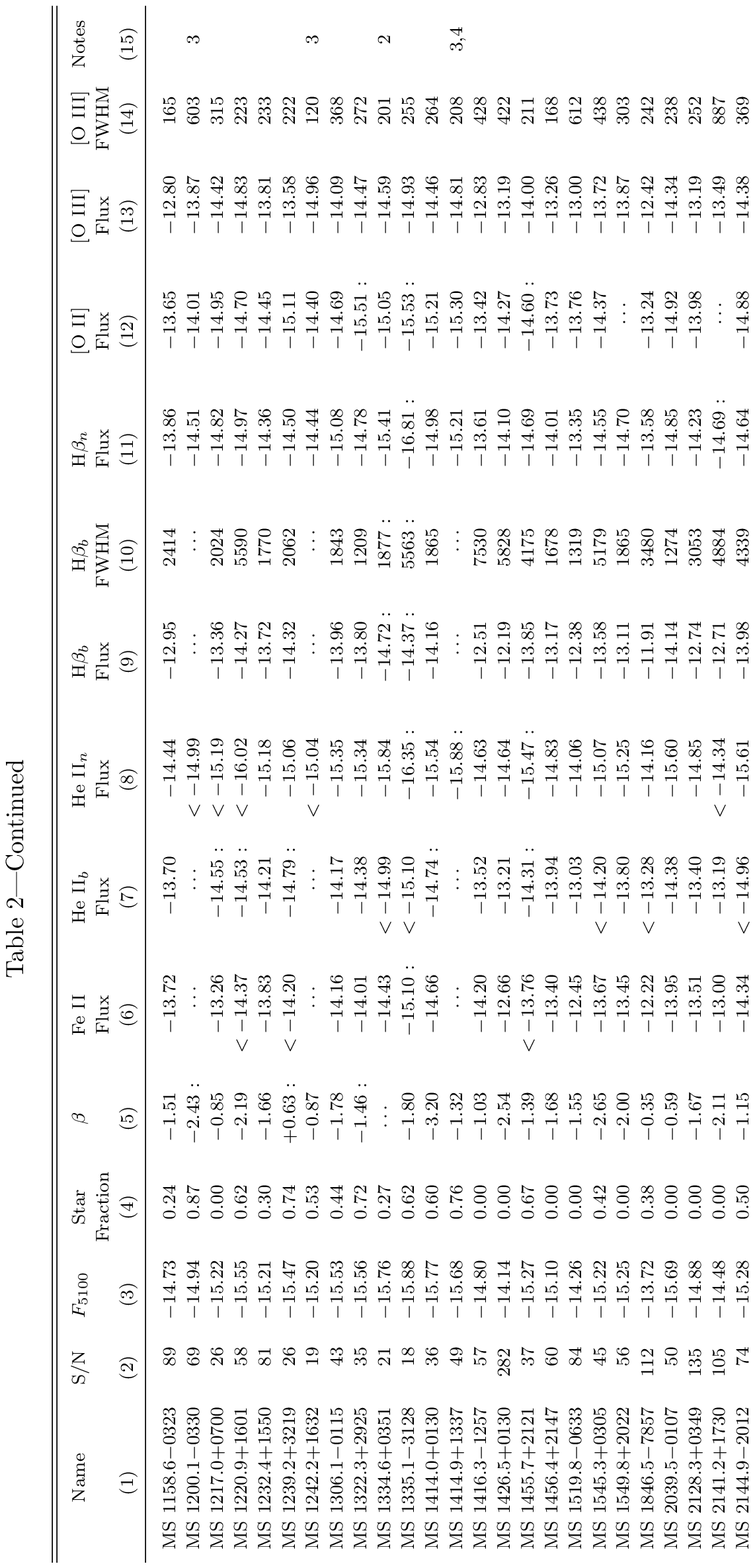,width=7.8in,angle=0}
\clearpage

\hskip -4in
\psfig{file=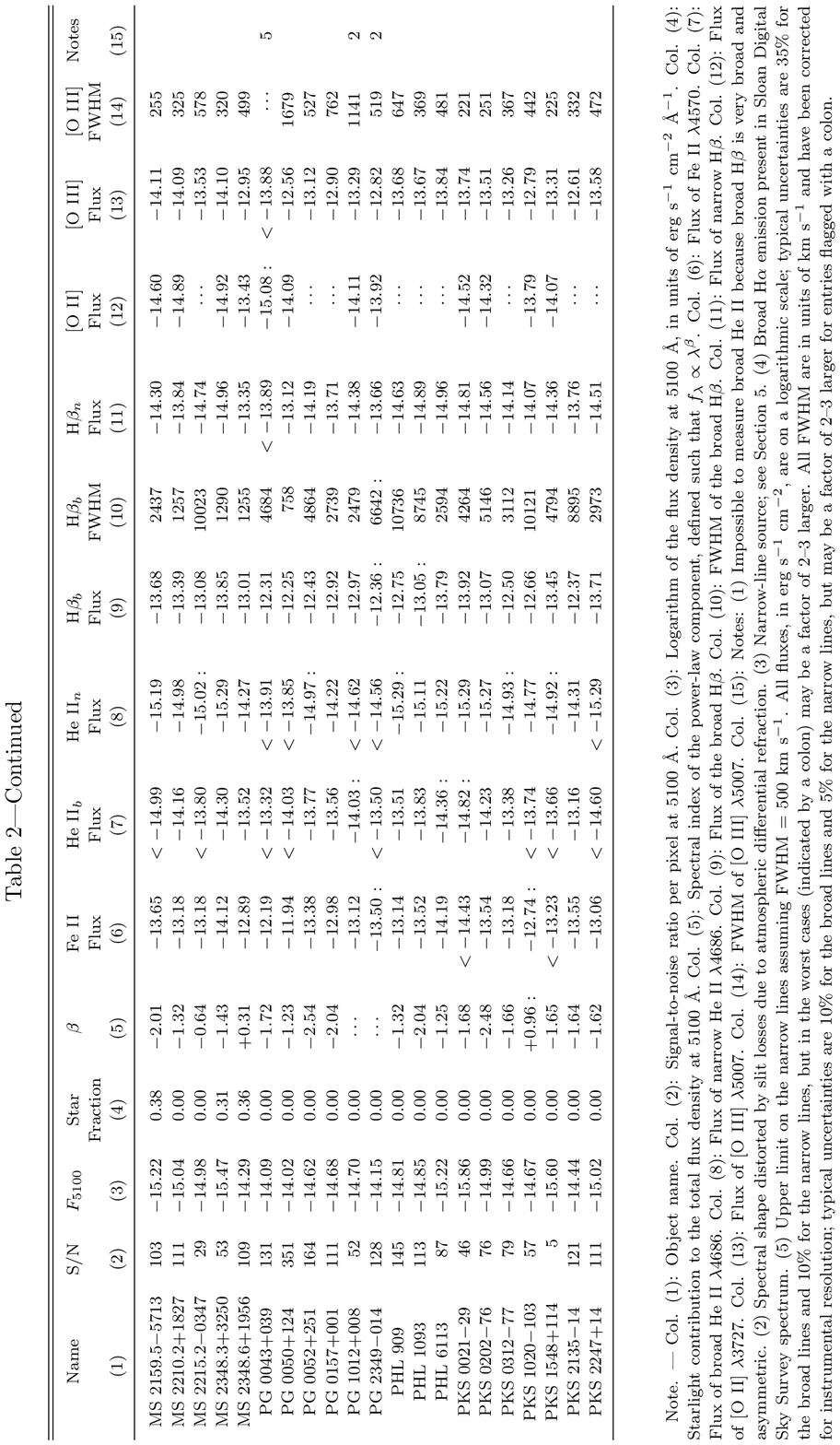,width=7.8in,angle=0}
\clearpage

\section{Measurements}

\subsection{Spectral Decomposition}

Although the primary purpose of this paper is to present the spectral atlas
of our database, we also measure a number of basic spectral parameters for the
continuum and strong emission lines commonly used by the AGN community.  Our 
own forthcoming host galaxy analyses will draw heavily from this database.
These measurements are summarized in Table~2.  Because of our wavelength 
coverage, we concentrate on the following emission lines: \oii\ \lamb3727, 
\feii\ \lamb4570, \heii\ \lamb4686, \hb\ \lamb4861, and \oiii\ \lamb\lamb4959, 
5007.  Our approach closely follows that of Greene \& Ho (2005b) and Kim et 
al. (2006), which the reader can consult for more details.  Here we briefly 
summarize a few key points.

The optical continuum is a complex mixture of several components, which must 
be modeled and subtracted prior to measuring the emission lines.  We decompose 
the continuum using a model consisting of up to three components: (1) starlight
from the host galaxy, (2) featureless continuum from the AGN, and (3) \feii\ 
emission.  We do not account for internal extinction, as there is no 
unambiguous, universally accepted method of doing so this type of data.  We 
include a galaxy component only if the observed spectrum contains a 
sufficiently strong starlight component.  Following Kim et al. (2006), we 
require that the equivalent width of the \caii\ K line exceed 1.5 \AA; this 
corresponds roughly to a starlight contribution of \gax 10\% to the local 
continuum.  The galaxy continuum is modeled by a linear combination of a 
G-type and a K-type giant, which, in most cases, suffices to match host 
galaxies with a predominantly evolved stellar population.  Some sources show 
a significant contribution from intermediate-age stars, as evidenced by the 
presence of strong higher-order Balmer lines (e.g., MS~1220.9$+$1601, 
MS~2144.9$-$2012, and MS~2159.5$-$5713; see Figure~1).  For these cases adding 
an additional A-type star does an adequate job of representing the young 
population.  Over the limited wavelength range under consideration, the 
featureless nonstellar continuum can be approximated using a single power-law 
function.  In a few cases the AGN continuum is slightly more complex, and 
a better fit can be achieved using the sum of two power-law functions.  
Finally, blends of broad \feii\ transitions form a complicated pseudocontinuum 
that affects significant portions of the ultraviolet and optical spectrum of 
most type~1 AGNs.  Following standard practice, we model the \feii\ component 
using a scaled and broadened Fe template derived from observations of the 
narrow-line Seyfert 1 (NLS1) galaxy I~Zw~1 (Boroson \& Green 1992); the
Fe template was kindly provided by T. Boroson.  As in Hu 
et al.  (2008a, 2008b), we allow both the width and the radial velocity of the 
Fe template to be free parameters in the fit, because the kinematics of \feii\ 
need not be identical to those of broad \hb.  The Boroson \& Green Fe 
template, unfortunately, does not extend below 3700 \AA.  Because of this, we 
do not bother to include the Balmer continuum in the continuum model, since 
the fit below 3700 \AA\ should not be trusted anyway.

The fit is performed over the following spectral regions devoid of strong 
emission lines: $3900-3950$, $4020-4070$, $4170-4260$, $4430-4570$, 
$5100-5170$, $5210-5500$, and $6150-6200$ \AA.  In practice, however, we 
adjust the exact fitting regions to achieve the best fit.  Figures 2 and 3 
give examples of two sources for which the AGN is sufficiently strong that the 
continuum can be modeled with just the power-law and Fe components; in the 
case shown in Figure~4, the stellar component clearly must be included too.

After continuum subtraction, we fit the residual spectrum in order to measure 
the parameters of several prominent emission lines.  For the narrow emission 
lines, we follow the same procedure used by Kim et al. (2006).  If the \sii\ 
\lamb\lamb6716, 6731 doublet is included in the bandpass and the S/N is 
adequate, we use the profile of \sii\ to constraint \nii\ \lamb\lamb6548, 6583 
and the the narrow component of \hal\ and \hb.  However, in the majority of 
the objects \sii\ lies outside of our spectral coverage.  In this situation we 
have no choice but to rely on \oiii\ as a template for the narrow lines.  As 
in Greene \& Ho (2005a), we fit each of the lines of the \oiii\ \lamb\lamb4959, 
5007 doublet with one or two Gaussians, depending on whether they show an 
extended or asymmetric wing.  With the profile of the narrow component thus 
constrained, we fit the broad component of the permitted lines using as many 
Gaussian components as necessary to achieve an acceptable fit (typically only 
2--3 suffice).

The \heii\ \lamb4686 emission line, while diagnostically important, is 
challenging to measure accurately because it is much weaker than \hb\ and 
because it lies on the shoulder of the strong \feii\ \lamb4570 blend to the 
blue and \hb\ to the red.  We treat \heii\ in the same manner as \hb, using 
\oiii\ as a template for its narrow component, and a multi-Gaussian model for 
the broad component.  If either component is undetected, we set an upper limit 
based on 3 times the rms noise of the local continuum and an assumed velocity 
width; for the narrow component, we use the \oiii\ profile, whereas the broad 
\hb\ profile is used for the broad component.

As in previous studies (e.g., Boroson \& Green 1992; Marziani et al. 2003), we 
select the prominent \feii\ blend at 4570 \AA\ to represent the strength of 
the optical Fe emission.  The flux of the feature is integrated over the 
region 4434--4684 \AA.  If undetected, we calculate the 3 $\sigma$ upper 
limit from the rms noise over this region.

Finally, we give measurements for \oii\ \lamb3727, whose strength provides
constraints on the ongoing star formation rate in AGNs (Ho 2005).  As in 
Kim et al. (2006; see also Kuraszkiewicz et al. 2000), we measure the line 
strength by simply fitting a single 
Gaussian with respect to the local continuum near 3700 \AA.  This procedure 
suffices because the \oii\ doublet remains unresolved at the relatively low 
resolution of our observations, and measuring the line locally bypasses the 
complications of the poor continuum fits in the blue part of the spectra. 
Upper limits are set using the local rms noise and the velocity width of \oiii.

The velocity widths of \oiii\ listed in Table~2 pertain to the FWHM of the 
final model for the entire line profile, and have been corrected for 
instrumental resolution by subtracting it in quadrature.

\subsection{Uncertainties}

Robust uncertainties are notoriously difficult to derive for spectral 
measurements of the type presented above.  In almost all cases the 
formal error bars from the fits underestimate the true errors, which 
are dominated by systematic uncertainties in the myriad assumptions that 
enter into the complicated fits.  For this reason we resist assigning 
specific error bars to the entries in Table~2.  Nevertheless, based on 
past experience with analysis of this type (e.g., Ho et al. 1997; Greene \& Ho 
2005b; Kim et al. 2006; Hu et al. 2008a, 2008b), in the notes to Table~2 
we give some rough estimates of the typical uncertainties involved.  

\vskip 1.0cm
\section{General Properties of the EMSS AGNs}

As the EMSS AGNs have never been analyzed spectroscopically in a quantitative
manner, and our survey contains a significant fraction (80\%) of the sample 
studied by Schade et al. (2000), we will give a few general statistics on the 
spectroscopic properties of these objects.  The EMSS was conducted in the 
0.3--3.5 keV band and, as such, is expected to be biased toward sources that 
are bright in soft X-rays.  Because of their strong soft X-ray emission 
(e.g., Boller et al. 1996; Grupe et al. 2004), NLS1s should be overrepresented 
in AGNs selected on the basis of their soft X-ray emission compared to 
selection in other wavelengths (e.g., optical).  This is especially true 
because of the relatively low luminosities probed by the EMSS. We confirm this 
expectation.  Among the 51 EMSS sources with detected broad H$\beta$ emission, 
33\% have H$\beta$ FWHM $<$ 2000 \kms, the conventional linewidth criterion 
for NLS1s.  This fraction increases to 47\% if we relax the (somewhat 
arbitrary) FWHM cutoff to 2500 \kms.  As expected from previous studies, these 
sources tend to show relatively weak \oiii\ lines but prominent \feii\ 
emission.  By comparison, within the sample of $\sim 8500$ low-redshift ($z$ 
\lax 0.35), optically selected type~1 AGNs studied by Greene \& Ho (2007), the 
fraction of sources with H$\alpha$ FWHM $<$ 2000 \kms\ is 24\%, increasing to 
40\% for FWHM $<$ 2500 \kms.  [To zeroth order, broad H$\alpha$ and H$\beta$ 
have similar line profiles (Greene \& Ho 2005b).]

In terms of their luminosities, all of the EMSS sources should be regarded as 
Seyferts rather than quasars.  Their broad H$\beta$ luminosities range from 
$\sim 10^{39}$ to $10^{43}$ \lum, with a median value of $\sim 10^{42}$ \lum; 
using a standard conversion from line to continuum luminosity (Greene \& Ho 
2005b), this corresponds to a median absolute magnitude of only $M_B \approx 
-19$, roughly near the knee of the luminosity function of local Seyfert 
galaxies (see Figure 8 in Ho 2008).  Despite their modest luminosities, as 
discussed in M. Kim et al. (in preparation), the EMSS sources are radiating at 
a healthy fraction of their Eddington rates ($L_{\rm bol}/L_{\rm Edd} \approx 
0.01-1$) because their black hole masses are relatively low ($M_{\rm BH} 
\approx 10^6-10^8$ \solmass), consistent with their hosts being mostly 
spiral galaxies (Schade et al. 2000).

While our original intent was to observe type~1 AGNs for which we can estimate 
black hole masses, 10 objects from the EMSS sample turn out to reveal only 
narrow emission lines in our spectral range (see Table~2).  Of these, two 
have Sloan Digital Sky Survey spectra that extend further to the red than our 
spectra.  MS~1058.8$+$1003 shows a double-peaked broad H$\alpha$ line, and the 
relative intensities of its narrow lines qualify it as a low-ionization 
nuclear emission-line region; following the nomenclature of Ho et al. (1997), 
it should be classified as a LINER 1.9.  MS~1414.9$+$1337 also has weak 
broad H$\alpha$ emission, but its higher-ionization narrow-line spectrum 
qualifies it as a Seyfert 1.9.   As for the rest, only three (MS~0039.0$-$0145,
0516.6$-$4609, and 1110.3$+$2210) appear to be genuine type~2 Seyferts, as 
judged by their large \oiii/H$\beta$ ratios and, for the latter two, detection 
of relatively strong \oi\ \lamb6300.  Without the help of the diagnostic lines 
near H$\alpha$ (see Ho 2008), the physical nature of the remaining five 
objects --- MS~0944.1$+$1333,  1108.3$+$3530,  1114.4$+$1801,  1200.1$-$0330, 
and 1242.2$+$1632 --- is somewhat ambiguous.  But judging from the relative 
strengths of \oii, \oiii, and H$\beta$, we suspect that these sources 
are not powered by AGNs but rather by star formation.

\vskip 1.0cm
\section{Summary}

We have used the Magellan 6.5~m Clay Telescope to obtain moderate-resolution 
($R \approx 1200-1600$) optical spectra covering  the rest-frame region 
$\sim 3600-6000$ \AA\ for a sample of 94 low-redshift ($z$ \lax\ 0.5) type~1 
AGNs.  Although some of the objects are well-known sources, the majority do 
not have reliable spectroscopy in the literature that can be used to estimate 
black hole masses from their broad emission lines.  We pay special attention
to the sample of soft X-ray-selected (EMSS) AGNs with good-quality {\it HST}\ 
images studied by Schade et al. (2000).  Eight of the sources turn 
out to be narrow-line objects that were previously misclassified as broad-line
AGNs; of these only three are Seyfert~2 galaxies and the rest appear to be 
powered by stars.  We present a spectral atlas of our sample and basic 
measurements for a number of prominent, commonly used emission lines.  These 
data will be used in a separate paper aimed at studying the relationship 
between black hole mass and the properties of their host galaxies.  

\acknowledgements
This work was supported by the Carnegie Institution for Science and by
NASA grants HST-AR-10969.03-A and HST-GO-10428.11-A from the Space Telescope
Science Institute (operated by AURA, Inc., under NASA contract NAS5-26555).
M.~K. was supported by the Korea Science and Engineering Foundation (KOSEF) 
grant No. 2009-0063616, funded by the Korean government (MEST).  We made use 
of the NASA/IPAC Extragalactic Database (NED), which is operated by the Jet 
Propulsion Laboratory, California Institute of Technology, under contract with 
NASA.  We thank Todd Boroson for making available his Fe template spectrum.
We are grateful to an anonymous referee for a constructive review of this 
work.

\clearpage

%%REFERENCES
%\clearpage

\end{document}